\documentclass[aps,prd,reprint,a4paper,showpacs,showkeys,amsfonts,amssymb,amsmath,nofootinbib,floatfix]{revtex4-1}
\usepackage[T1]{fontenc}
\usepackage{textcomp}
\usepackage{txfonts}

\usepackage{color}
\usepackage{graphicx}
\usepackage{hyperref}
\hypersetup{colorlinks=true}
\hypersetup{urlcolor=black}
\hypersetup{linkcolor=blue}
\hypersetup{citecolor=blue}

\newcommand{\phpr}{\hphantom{'}}

\newcommand{\ppnn}{\pi^0\pi^0 \,H\to \pi^0\pi^0}
\newcommand{\ppnc}{\pi^0\pi^0 \,H\to \pi^+\pi^-}
\newcommand{\abs}[1]{\lvert #1 \rvert}
\newcommand{\uni}[1]{\hat{#1}}
\newcommand{\na}{\nabla}
\newcommand{\nn}{\nonumber}
\newcommand{\ga}{\gamma}

\newcommand{\la}{\langle}
\newcommand{\ra}{\rangle}
\newcommand{\vp}{\mathbf{p}}
\newcommand{\vq}{\mathbf{q}}
\newcommand{\vv}{\mathbf{v}}

\begin{document}

\title{On the size of the $\sigma$ meson and its nature}
\date{\today}
\author{M.~Albaladejo}
\email{albaladejo@um.es}
\author{J.~A.~Oller}
\email{oller@um.es}
\affiliation{Departamento de F\'{\i}sica. Universidad de Murcia, E-30071 Murcia, Spain.}

\begin{abstract}

In this work the nature of the $\sigma$ or $f_0(600)$ resonance is discussed by evaluating its quadratic scalar radius, $\la r^2\ra_s^\sigma$. This allows one to have a quantitative estimate for the size of this resonance. We obtain that the $\sigma$ resonance is  a compact object with $\la r^2\ra_s^\sigma=(0.19\pm 0.02)-i\,(0.06\pm 0.02)$ fm$^2$. Within our approach, employing unitary chiral perturbation theory, the $\sigma$ is a dynamically generated resonance that stems from the pion-pion interactions. Given its small size we conclude that the two pions inside the resonance are merged. A four-quark picture is then  more appropriate. However, when the pion mass increases, for pion masses somewhat above 400~MeV, the picture of a two-pion molecule is the appropriate one. The $\sigma$ is then a spread $\pi\pi$ bound state.  These results are connected with other recent works that support  a non standard nature of the $\sigma$ as well, while fulfilling strong QCD constraints, as well as with lattice QCD.

 We offer a detailed study of the 
low-energy $S$-wave $\pi\pi$ scattering data from where we extract our values for the threshold parameters of $S$-wave  $\pi\pi$ phase shifts,  the ${\cal O}(p^4)$ chiral perturbation theory low energy constants as well as the $\sigma$ pole position. From the comparison with other accurate determinations in the literature we obtain the average values for the isospin 0 $S$-wave $\pi\pi$ threshold parameters, $a_0^0=0.220\pm 0.003$, $b_0^0=0.279\pm 0.003$~$M_\pi^{-2}$, and for the real and imaginary parts of the  $\sigma$ pole position in $\sqrt{s}$,  $458 \pm 14-i\,261\pm 17$~MeV. The quark mass dependence of the size of the $\sigma$, its mass and width are considered too. The latter agree accurately  with a previous lattice QCD calculation. The fact that the mass of this resonance tends to follow the threshold of two pions is a clear indication  that the $\sigma$ is a dynamically generated meson-meson resonance.

\end{abstract} 

\maketitle


\section{Introduction}\label{sec.intro}

The lightest resonance in QCD with the quantum numbers of the vacuum, $J^{PC}=0^{++}$, is the $\sigma$ or $f_0(600)$ resonance \cite{pdg}. Its connection with chiral symmetry has been stressed since the sixties in the linear sigma model \cite{levi}, while its tight relation with the non-linear sigma model was realized in the nineties. In this respect there have been several papers that clearly connect this resonance with  chiral dynamics of the two-pion system. One has first to mention the works of Truong and collaborators \cite{tr1,e3pi.tr,prl.tr,dobado} who first emphasized the important role played by the null isospin ($I$)  $S$-wave $\pi\pi$ final state interactions in several processes giving rise to a strong numerical impact on the estimations based on current algebra technique or chiral perturbation theory (ChPT) \cite{Gasser:1983yg,chptsu3,ulf,pich,bernard}. A notoriously improved comparison with experiment was then obtained, \textit{e.g.} for $K_{\ell_4}$ decays \cite{tr1}, $\eta\to 3\pi$ \cite{e3pi.tr}, scalar and pion vector form factors \cite{prl.tr} and $\pi\pi$ scattering \cite{dobado}. These works stress the role of the right-hand or unitarity cut and make use of a method to resum unitarity based on the expansion of the inverse of a form factor or scattering amplitude. This is the so called Inverse Amplitude Method (IAM), that in the end is analogous to a Padd\`e  method of resummation. Within this technique the $\sigma$ pole was first obtained in Ref.~\cite{pela1}, together with the $K^*$ and $\rho$ resonances in the $P$-waves. However, due to the lack of coupled channels, no further light scalar resonances were generated, in particular the  $f_0(980)$ and  $a_0(980)$.

Independently, the $\sigma$ resonance pole was also obtained simultaneously in Ref.~\cite{npa}, together with the $I=0$ $f_0(980)$ and $a_0(980)$ resonances. The associated  amplitudes were determined by solving the Bethe-Salpeter equation taking as potential the lowest order ChPT Lagrangian. Only one free parameter (a natural sized cut-off)  was involved. Later on, when the IAM was extended to coupled channels \cite{prl}, it was possible to obtain in Refs.~\cite{prl,prdlong,guerrero} the $\sigma$, $f_0(980)$, $a_0(980)$ and $\kappa$ resonances altogether, that is, the whole nonet of the lightest scalar resonances \cite{jaffe1,beveren,nd,mixing,black,closetorn}, together with the nonet of the lightest vector resonances.

The approach of Ref.~\cite{npa}, based on solving a Bethe-Salpeter equation, was put on more general grounds in Ref.~\cite{nd} by applying the N/D method \cite{mandelstam}. In this way, it is possible to include higher orders in the chiral counting \cite{plb,higgs} as well as explicit resonant fields \cite{colla}, if required. Later works based on this scheme are Refs.~\cite{jamin,Albaladejo,procalba,pseuscalar}. With this approach \cite{nd} one builds a unitarized meson-meson scattering amplitude by solving the N/D equation in an algebraic way so that an approximate solution is obtained by treating perturbatively  the crossed cuts. As a result,  the ChPT expansion is reproduced order by order, while the  unitarity cut is resummed \cite{plb}. In this respect, one should stress that the crossed cuts can be treated perturbatively for the isoscalar $\pi\pi$ $S$-wave. Its size was estimated to be smaller than 10\% in Ref.~\cite{nd} along the physical region for energies up to around 1~GeV. Indeed, different approaches with various degrees of sophistication provide very similar values for the $\sigma$ pole resonance parameters, mass and width. Either by employing just the leading order (LO) ChPT \cite{npa} (without left-hand cut at all), next-to-leading order (NLO) \cite{pela1} or next-to-next-to-leading order (N$^2$LO) \cite{hannah}. In these two later references the left-hand cut is included as calculated by ChPT at one and two-loop orders, respectively. The fact that the results are very similar clearly indicates that the left-hand cut is indeed a perturbation. The $\sigma$ pole positions in $\sqrt{s}$, with $s$ the total center of mass energy 
squared, obtained in these works  are: $\sqrt{s_\sigma}=468-i~194$~MeV \cite{npa}, $440-i~245$~MeV \cite{pela1} and $445-i~235$~MeV \cite{hannah}. In the following we identify the mass and half width of the $\sigma$ resonance from the pole position as $M_\sigma-i\,\Gamma_\sigma/2\equiv \sqrt{s_\sigma}$. 

More recently, Ref.~\cite{caprini}, based on the solution of the Roy equations \cite{roy} and ChPT at two-loops \cite{cola7,leutyplb}, obtained the value 
$445^{+16}_{-8}-i~272^{+9}_{-13}$~MeV. The Roy equations implement crossing symmetry exactly, while the previous references \cite{npa,pela1,hannah,nd} do it perturbatively. The fact that all these pole positions for the $\sigma$ lie rather close to each other (particularly  one can say that convergence is reached very accurately for the real part) is another indication for the correctness of treating crossed-channel dynamics perturbatively, as done in the framework of Refs.~\cite{nd,nn} (see also \cite{Albaladejo:2011bu,Albaladejo:2012sa,arriola1}). Indeed, to our mind, both schemes are complementary because the Roy equations  need for their implementation of the knowledge of large amount of data in several partial waves up to high energies, which is affected by systematics errors in experiments (many of them old ones) and also in theory (\textit{e.g.} high energy extrapolations), not always well under control. Let us also mention that all these analyses neglect altogether the inelasticity due to the $4\pi$ channel in $\pi\pi$ $S$-waves so that,  up to the opening of the $K\bar{K}$ threshold at around 1~GeV, no inelasticity is assumed. The $4\pi$ channel was approached in Ref.~\cite{Albaladejo} as $\sigma\sigma$ and $\rho\rho$ states (with their couplings to all the channels predicted from chiral dynamics) and found the $\sigma$ pole at $456\pm6-i~241\pm7$~MeV.\footnote{In addition this reference was able to reproduce simultaneously all the isoscalar $S$-wave resonances  quoted in the PDG \cite{pdg} from $\pi\pi$ threshold up to 2~GeV. A coherent picture of the scalar sector dynamics and spectroscopy then arose, including the identification of the lightest scalar glueball.} This pole position is quite close to those in the previous references and compatible with the result $484\pm 17-i~255\pm 10$~MeV from Ref.~\cite{martin}. Thus, since the pole positions of Refs.~\cite{npa,pela1,hannah,caprini,Albaladejo,martin} lie so close to each other we could conclude that our present knowledge on the pole position of the $\sigma$ resonance is quite precise and, furthermore, we understand the underlying physics at the hadronic level.

Between  earlier approaches to the previous discussed results based on ChPT concerning the lightest scalars, we have 
Refs.~\cite{jaffe1,jaffe2} within the MIT bag model that already in the late seventies predicted a complete nonet of four-quark $0^{++}$ resonances (comprising the $\sigma$, $f_0(980)$, $a_0(980)$ and $\kappa$), with  $M_\sigma=660\pm 100$~MeV and $\Gamma_\sigma=640\pm 140$~MeV.  The four-quark nature of the lightest scalars is also favored in Refs.~\cite{acha4,acha3,acha2,acha1} attending to scattering and production data, including two-photon fusion, $J/\Psi$ and $\phi$ decays, and in Refs.~\cite{polosa,zhou2010}. The important role played by two-meson unitarity for understanding the scalar sector for $\sqrt{s}\lesssim 1$~GeV was also stressed in Ref.~\cite{beveren} (a similar approach was later followed in Ref.~\cite{roos}), employing a unitarized chiral quark model, and in Ref.~\cite{speth}, within the J\"ulich meson-exchange models. Considerations based on increasing the QCD number of colors, $N_C$,  were exploited in Refs.~\cite{nd,pelanc,rios,zhou2010,guo11prd}, showing that the $\sigma$ resonance has a non-standard $N_C$ dependence. This can be done more safely  for $N_C\gtrsim 3$, not too large, while statements for $N_C\gg 3$ depend much more on fine details of the approach \cite{chino,guo,guo2,arriola1,arriola2,guo11prd,nievepich}. QCD sum rules were also applied for the study of the lightest scalar meson, \textit{e.g.} in Refs.~\cite{narison23,narison22,qcdssr.steele,qcdssr.toki,steele}. It is argued too that the $\sigma$ resonance is the chiral partner of the pion \cite{hatsuda,bernard2,kyoto,hatsuda2} and the way in which the $\sigma$ pole evolves when approaching  the chiral symmetry restoration limit is different according to the nature of this resonance \cite{hyodo}.

From an experimental point of view new interest is triggered on the $\sigma$ resonance from recent high-statistics results, \textit{e.g.} $J/\Psi\to \omega\pi\pi$ where a  conspicuous peak is seen \cite{besii}. Indeed, this decay mode was the  first clear experimental signal of a $\sigma$ resonance \cite{dm2,mark3}. Another marked peak around the $\sigma$ energy region is also observed in several heavy meson decays. \textit{E.g.} it was observed with high statistical significance in $D\to \pi^+\pi^-\pi^+$ \cite{e791}. Both types of decays present a strong peak in the $\sigma$ mass energy region because the absence of the Adler zero in the pion scalar form factor, as explained in Refs.~\cite{ddecays,bugd}.\footnote{One can explain consistently both types of decays in terms of the pion scalar form factors \cite{fioller,ddecays,lahde}} Another field of increasing activity, both experimental \cite{belle,frascatti} and theoretical, concerns the fusion of two photons into a pair of pions and from there to extract the width of the $\sigma$ to $\gamma\gamma$ \cite{penfoton,ollerf1,gamma,narison,besojudas}. This is also expected to shed light on the nature of the $\sigma$ meson \cite{penfoton}. 

The relative strength of the $\sigma$ coupling to $K\bar{K}$ compared to $ \pi\pi$ is also taken as an important property in order to disentangle between different models for the nature of the $\sigma$ meson ($q\bar{q}$, four-quarks, glueball or $\pi\pi$-molecule), as stressed in Ref.~\cite{narison22}. This reference points out that the not so much suppressed  coupling of the $\sigma$ to $K^+K^-$  ($g_{\sigma K^+K^-}$), as compared with that to $\pi^+\pi^-$ ($g_{\sigma \pi^+\pi^-}$), $|g_{\sigma K^+K^-}|/|g_{\sigma \pi^+\pi^-}|=0.37\pm 0.06$ \cite{narison22}, is a key ingredient to advocate for a gluonium nature of the $\sigma$ meson. According to Ref.~\cite{narison22}, a simple $q\bar{q}$ interpretation of the $\sigma$ fails to explain the large width of the $\sigma$ while a four-quark scenario has difficulties to explain its large coupling to $K^+K^-$.  It is then worth emphasizing that the $T$-matrices obtained in Refs.~\cite{npa,nd,Albaladejo}  also predict a ratio for the $\sigma$ couplings to $K^+K^-$ and $\pi^+\pi^-$ in perfect agreement with the value above of Refs.~\cite{narison22,narison23,narison24}. Explicitly, we have $|g_{\sigma K^+ K^-}|/|g_{\sigma\pi^+\pi^-}|=0.36\pm 0.04$ from the average collected in Ref.~\cite{mixing}. However, in our case this stems from the dynamical generation of the $\sigma$ resonance from the Goldstone boson dynamics associated to the strong scalar isoscalar  $\pi\pi$ interaction.  We also stress that this approach has been confronted with a large amount of data from different reactions, both scattering and production experiments, in most of the reactions already quoted in this introduction.

 One of the aims of this work is to show that the often identification of dynamically generated resonances from the interactions of two mesons (pions in our case) as  meson-meson molecules is misleading. As we show here, depending on the meson mass, one can have situations where the size of a dynamically generated meson-meson resonance is certainly too small to be qualified as a two-meson molecule. Indeed, its size could be as small as that of one of the mesons involved in their formation. The fact that the $\sigma$ is such a tight compact object clearly hints that the two pions pack so much that it is not meaningful anymore to keep their identities separately. At this stage, a four quark compact resonance seems a more appropriate picture. This is also supported by the $N_C$ evolution of the $\sigma$-pole trajectory which is clearly at odds with the expectations for a purely $\bar{q}q$ or glueball resonance, but in the lines of what it is expected for a meson-meson or four quark resonance \cite{manohar,chino,guo,guo2,arriola1,arriola2,guo11prd,nievepich}. However, by increasing the pion mass the $\sigma$ resonance pole tends to follow the two pion threshold, and when it is close to the latter its size increases, becoming a spread object. This is a clear indication for the molecular character of the $\sigma$ for large enough pion 
masses, $M_\pi \gtrsim 400\ \text{MeV}$. In addition, let us also emphasize that our work is the first calculation of the size of the $\sigma$ resonance. This is a novel way to study its nature in the literature.

The rest of the paper is organized as follows. In Sec.~\ref{sec:chilag} we give a short introduction to the $SU(2)$ ChPT Lagrangians both at LO and NLO that are used in the rest of the paper. Next we dedicate Sec.~\ref{sec:pipiscattering} to evaluate $\pi\pi$ scattering at one-loop order in  Unitary ChPT. A wide set of data is fitted, including some recent lattice QCD determinations as a function of $M_\pi$.  We pay special attention to the threshold parameters and the $\sigma$ pole position. For these quantities we also compare with previous phenomenological determinations and the lattice QCD results on the dependence of the $\sigma$ pole mass as a function of the pion mass. We dedicate Sec.~\ref{sec:sigmaff} to the calculation of the scalar form factor of the $\sigma$ resonance. First pion scattering in the presence of a scalar source is discussed. The scalar form factor of the $\sigma$ is calculated from the double $\sigma$ pole present in the amplitude for the previous process, once $\pi\pi$ initial and final state interactions are taken into account. Then, we determine the quadratic scalar radius of the $\sigma$ and then have some information on the size of this resonance. We stress that this radius is pretty small, around 0.5~fm indicating that the $\sigma$ is a compact object. We also discuss the relation between the value of the $\sigma$ scalar form factor at the origin and the dependence of the $ \sigma$ pole with the pion mass, related by the Feynman-Hellmann theorem. Both issues, the quadratic scalar radius and the Feynman-Hellman theorem, are addressed in Sec.~\ref{sec:r2FH}. After concluding in Sec.~\ref{sec:conclusions}, we dedicate Appendix~\ref{app:loop} to the loop functions used throughout the amplitudes calculated, which are in turn given in Appendix~\ref{app:amp} for pion scattering in the presence of a scalar source.

\section{$SU(2)$ Chiral Lagrangians}
\label{sec:chilag}

We follow the standard ChPT counting and the processes under consideration, the scattering of pions with and without the presence of a $c$-number external scalar source, are calculated both at LO and NLO. The  chiral power counting of a connected diagram, $p^D$ (where $p$ is a generic small momentum compared to $\Lambda_{\textrm{ChPT}}\simeq 1$~GeV), obeys the equation \cite{Gasser:1983yg,pich}
\begin{align}
D&=2+\sum_d N_d(d-2)+2L~.
\end{align}
In this equation, $d$ is the chiral dimension of a vertex, $N_d$ the number of vertices with dimension $d$ and $L$ is the number of loops. Each derivative increases the counting by one unit and the lightest quark masses add two units to $D$. The LO calculation has $D=2$ with no loops ($L=0$) and involves only  $d=2$ vertices. For the NLO, $D=4$, and one has diagrams with $L=1$ that involve only $d=2$ vertices. There are  also diagrams with $L=0$  with only one $d=4$ vertex, with the rest of vertices having $d=2$.

Up to NLO, ${\cal O}(p^4)$, one has to consider the $SU(2)$ chiral Lagrangians at $\mathcal{O}(p^2)$, $\mathcal{L}_2$,  and $\mathcal{O}(p^4)$, $\mathcal{L}_4$, that we take from Ref.~\cite{Gasser:1983yg}:
\begin{eqnarray}
\mathcal{L}_2 & = & \frac{F^2}{2} \left( \na_{\mu} U^{\rm{T}} \na^{\mu} U + 2(\chi^{\rm{T}} U) \right) \label{eq:lagGL}~, \\
\mathcal{L}_4 & = & 
l_1 \left( \nabla^{\mu}U^{\rm{T}} \na_{\mu}U \right)^2
 +  l_2 \left( \nabla^{\mu}U^{\rm{T}} \na^{\nu}U \right) \left(\nabla_{\mu}U^{\rm{T}} \na_{\nu}U \right)  \nn \\
&+ & l_3 \left( \chi^{\rm{T}} U \right)^2 
 + l_4 \left( \nabla^{\mu}\chi^{\rm{T}} \na_{\mu}U \right) +\cdots
\label{L4_GL}
\end{eqnarray}
where $F$ is the pion weak-decay constant in the chiral limit, the terms proportional to the $l_i$ are the NLO chiral counterterms and the ellipsis indicate terms not shown because are not needed here. The pion fields are included through the $O(4)$ real vector field $U(x)$ of unit length, $U^T U=1$, as:
\begin{align}
U^{\rm{T}} & = \left( U^0(x) , \vec{U}(x) \right)~, \nn \\
\vec{U}(x) & = \frac{\vec{\pi}(x)}{F} = \frac{1}{F} \left( \pi^1,\pi^2,\pi^3 \right)~, \nn \\
U^0(x) & =  \sqrt{1-\vec{U}(x)^2} = 1-\frac{1}{2}\vec{U}^2 - \frac{1}{8} \vec{U}^4 - \cdots 
\label{eq:pifielddef}
\end{align}
 We also use the relation between the charged and Cartesian pion fields given by
\begin{equation}
\pi^{\pm} = \frac{\pi^1 \mp i\pi^2}{\sqrt{2}} \quad\text{,}\quad \pi^{0} = \pi^3~.
\label{pif.def}
\end{equation}
 The explicit chiral symmetry breaking due to the finite $u$ and $d$ quark masses enters through the vector-field $\chi^{T}(x) = 2B(\hat{m}+s(x),p^i(x))$. Here, $2B\hat{m} = M^2$ is the pion mass at leading chiral order\footnote{In our notation $M$ represents the pion mass at LO, \textit{i.e.}, the parameter that appears directly from the Lagrangian, while $M_{\pi}$ refers to the physical pion mass.} and $\hat{m}$ is the algebraic mean of the $u$ and $d$ quark masses (we consider exact isospin symmetry). The fields $s(x)$ and $p^i(x)$ refer to the scalar and pseudoscalar $c$-number external sources, in order. The covariant derivative $\nabla_\mu$ reduces in the problem that we are studying to the standard  derivative, $\nabla_\mu \to \partial_\mu$, since we do not consider here external vector nor axial-vector currents. Finally, the parameter $B$ is related to the value of the quark condensate in the chiral limit  $\la\bar{q}^iq^j\ra=-\delta^{ij}F^2 B$ \cite{Gasser:1983yg}.  

In the following we employ the finite and scale independent constants $\bar{l}_i$ defined by 
\begin{eqnarray}
l_i & = & l_i^r + \gamma_i\frac{R}{32\pi^2} ~,\nn\\
l_i^r(\mu) & = & \frac{\gamma_i}{32\pi^2} \left( \bar{l}_i + \log \frac{M^2}{\mu^2} \right) ~, \nn\\
R & = & \mu^{n-4} \left( \frac{2}{n-4} - \left( \log 4\pi +\Gamma'(1) + 1 \right) \right)~,\label{eq:lecsNLO}
\end{eqnarray}
so that $\bar{l}_i$ is, up to a numerical factor, the  renormalized coupling constant $l_i^r$ at the scale 
$\mu=M\simeq M_\rho$. 
In the chiral limit the $\bar{l}_i$ are not defined as they are then divergent quantities. The needed $\gamma_i$ coefficients are \cite{Gasser:1983yg}:
\begin{equation}
\ga_1 = \frac{1}{3}~\text{,}\quad 
\ga_2 = \frac{2}{3}~\text{,}\quad
\ga_3 =-\frac{1}{2}~\text{,}\quad
\ga_4 = 2~\text{.}
\end{equation}
The infinite quantity $R$ is cancelled with the infinities that originate from loops, see Appendix~\ref{app:loop}.

\begin{figure}
\centering
\includegraphics[width=.45\textwidth,keepaspectratio]{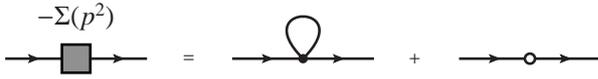}
\caption{Diagrams for the one-loop calculation of the pion self-energy. Full circles represent $\mathcal{O}(p^2)$ vertices, while the empty ones 
correspond to the $\mathcal{O}(p^4)$ vertices.\label{fig:pse}}
\end{figure}

The calculation of the pion self-energy, $-i\Sigma(p^2)$, is necessary in order to take into account the renormalization of the wave function of the initial and final pions. One has:
\begin{equation}
\label{sigmap2}
\Sigma(p^2) = \frac{3 M^2 A_0(M^2)}{2F^2} + \frac{2M^4 l_3}{F^2} - \frac{p^2 A_0(M^2)}{F^2}~\text{.}
\end{equation}
Notice that $\Sigma(p^2)$ is linear in its argument. The one-point function $A_0(M^2)$ is given in Eq.~\eqref{eq:loopA0}, Appendix~\ref{app:loop}, together with the different $n$-point loop function needed in this work.  We can write the self-energy Eq.~\eqref{sigmap2} as:
\begin{align}
\Sigma(p^2)&=\Sigma(M_\pi^2)+\Sigma'(M_\pi^2) (p^2-M_\pi^2)~.
\label{exp.linear}
\end{align}
The Dyson resummation gives for the renormalized propagator, $\Delta_R(p^2)$,
\begin{eqnarray}
i\Delta_R(p^2) = \frac{i}{(p^2 - M_\pi^2)(1-\Sigma'(M_\pi^2))} \equiv \frac{iZ}{p^2 - M_\pi^2}~,
\label{pp.ren}
\end{eqnarray}
where
\begin{align}
Z & \simeq  1+\delta Z=1 - \frac{A_0(M^2)}{F^2}~+{\cal O}(M_\pi^4)~\text{,}\nn\\
M_\pi^2  & = M^2 \left( 1 - \frac{M^2}{32\pi^2 F^2} \bar{l}_3 \right)+{\cal O}(M_\pi^6)~\text{.}
\label{mpi2}
\end{align}
Then, in order to take into account the renormalization of the pion wave function in our diagrams (both for $\pi\pi$ scattering and for the $\pi\pi s \to \pi\pi$ process), with four external legs, we have to multiply by a factor $(Z^{1/2})^4 = Z^2 = 1 + 2\delta Z+{\cal O}(M^4)$. In the following, we should keep in mind that the pion propagators employed are $i\Delta_R(p^2)$, Eq.~\eqref{pp.ren}, in terms of the physical pion mass. This will make simpler the calculation of some diagrams for the process $\pi\pi s \to \pi\pi$. Let us also mention that the amplitudes calculated are given in terms of the physical mass and weak decay constant of the pion. The latter is given by \cite{Gasser:1983yg}:
\begin{align}\label{eq:fpiNLO}
F_\pi & =  F\left( 1 + \frac{M^2}{16\pi^2 F^2}\ \bar{l}_4 \right)+{\cal O}(M_\pi^4)~.
\end{align}


\section{$\pi\pi$ scattering and the $\sigma$ meson}\label{sec:pipiscattering}

\subsection{The $\pi\pi \to \pi\pi$ amplitude}\label{subsec:pipiampli}

\begin{figure}\centering
\includegraphics[height=4cm,keepaspectratio]{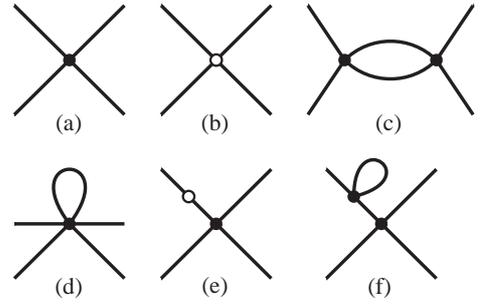}
\caption{Feynman diagrams for $\pi\pi$ scattering  up to NLO. Full circles represent $\mathcal{O}(p^2)$ vertices, while the empty ones 
correspond to the $\mathcal{O}(p^4)$ vertices.\label{fig:pipi}}
\end{figure}

The chiral Lagrangians exposed in Sec.~\ref{sec:chilag}  comprise four low energy constants (LECs), $\bar{l}_i$, at $\mathcal{O}(p^4)$. Additionally, our resummation procedure, explained below, includes a subtraction constant through the two-meson unitarity one-loop function. Before considering the $\pi\pi s \to \pi\pi$ amplitude, we must fix these free parameters. This is accomplished by comparing our results for the scalar $\pi\pi \to \pi\pi$ phase shifts with $I=0,~2$ with experiment, and also other observables with lattice QCD determinations.

We denote by  $\chi_n(s,t)$ the $I=0$ $\pi\pi$ scattering amplitudes calculated from Fig.~\ref{fig:pipi} in ChPT at ${\cal O}(p^n)$, with $n=2$ or 4. Their projection in $S$-wave are indicated by $\xi_n(s)$. Diagram a) is the LO contribution, while the rest of diagrams are the NLO ones. The last two diagrams, namely, e) and f) contribute to the wave-function renormalization of the pion external legs.  We introduce the usual Mandelstam variables $s$, $t$ and $u$. The variable $s$ corresponds to the total energy squared of the two pions in their center of mass frame (CM), while the other two are defined as:
\begin{align}
t & = -2 \vp^2(1-\cos\theta) \nn\\
u & = -2 \vp^2(1+\cos\theta) \nn\\
s + t + u & = 4M_\pi^2 \nn\\
\vp^2 & = \frac{s}{4} -M_\pi^2 \label{eq:stu}
\end{align}
Here, $\vp^2$ is the three-momentum squared of the pions in their CM and $\theta$ is the scattering angle in the same reference frame. The amplitudes $\xi_n(s)$ are then given by,
\begin{equation}\label{eq:norm_uni}
\xi_n(s) = \frac{1}{4} \int_{-1}^{-1}\!\!\! \text{d}\cos\theta\,\, \chi_n(s,t)~\text{.}
\end{equation}
In the previous equation an extra factor of $1/2$ has been included, in correspondence with the so called unitarity normalization \cite{npa}. The $I=0$ $\pi\pi$ state is symmetric under the exchange of the two pions so that the unitarity normalization avoids having to take into account the presence of the factor 1/2 whenever it appears as an intermediate state. In this way, the same formulas as for distinguishable particles can be employed. In the following of the paper we employ the unitarity normalization in all the isoscalar $\pi\pi$ matrix elements unless the opposite is stated.

Let us indicate by $T(s)$ the scalar-isoscalar unitarized $\pi\pi$ partial-wave amplitude. Following the unitarization method of Refs.~\cite{nd,plb}, the right-hand cut or unitarity cut is resummed by the master formula:
\begin{equation}\label{eq:tpipi_uni}
T(s) = \frac{V(s)}{1+V(s)G(s)}~\text{.}
\end{equation}
This formula is deduced by solving algebraically the N/D method \cite{mandelstam,nd}, treating perturbatively the crossed cuts, whereas the unitarity cut is resummed exactly. Here, $G(s)$ is the scalar two-point function, 
\begin{equation}
G(s) = \frac{1}{16\pi^2} \left( a + \log\frac{M_\pi^2}{\mu^2} - \sigma(s)\log\frac{\sigma(s)-1}{\sigma(s)+1} \right)~\text{,}
\end{equation}
with chiral order $p^0$. In the previous equation $\sigma(s) = \sqrt{1-4M_\pi^2/(s+i\epsilon)}$. The interaction kernel $V(s)$ has a chiral expansion, $V(s) = V_2(s) + V_4(s) + \cdots$, with the chiral order determined by the subscript. The different chiral orders of $V(s)$ are calculated by matching 
 $T(s)$ with its perturbative expansion calculated in ChPT. In this way up to ${\cal O}(p^4)$,
\begin{align}\label{eq:chiralexpansion}
T(s) &= \frac{V(s)}{1+ V(s)G(s)} \nn\\
&=\xi_2(s) + \xi_4(s) +\ldots \nn\\
&= V_2(s) + V_4(s) - V_2^2(s)G(s)+\ldots~,
\end{align}
where the ellipsis indicate ${\cal O}(p^6)$ and higher orders in the expansion. It results then:
\begin{align}
V_2(s) & = \xi_2(s)~, \nn\\
V_4(s) & = \xi_4(s) + \xi_2(s)^2 G(s)~.
 \label{eq:v4}
\end{align}

The finite piece of the unitarity term in Fig.~\ref{fig:pipi} (that is, the term of $\xi_4(s)$ that contains the unitarity cut and is proportional to the unitarity two-point one-loop function)  is given by:
\begin{equation}
\xi_4^U(s) = -\xi_2^2(s) \bar{B}_0(s)~.
\end{equation}
Here, $\bar{B}_0(s)$ is the two-meson loop in dimensional regularization, without the $R + \log(M^2/\mu^2)$ piece (that cancels out with the other infinite and scale dependent terms, see Eqs.~\eqref{eq:loopB0} and \eqref{eq:loopB0_sub} in Appendix~\ref{app:loop}). In this way, the kernel $V(s) = V_2(s) + V_4(s)$ has no unitarity cut because:
\begin{equation} 
\xi_4^U(s) + \xi_2^2(s)G(s) = -\xi_2^2(s)(\bar{B}_0(s) - G(s))~,
\end{equation}
and the cut cancels in the r.h.s. of the previous equation. The full unitarity cut arises from the denominator $1 + V(s)G(s)$  in Eq.~\eqref{eq:tpipi_uni}.

In this Section we have dealt with the $I=0$ unitarized amplitudes but, needless to say, the same formalism applies to the $I=2$ ones, by just changing the kernel $V(s)$. We additionally note here that the same subtraction constant is used for both channels, as required by isospin symmetry \cite{jido}.


\subsection{Fits and the $\sigma$ meson}
\begin{figure*}
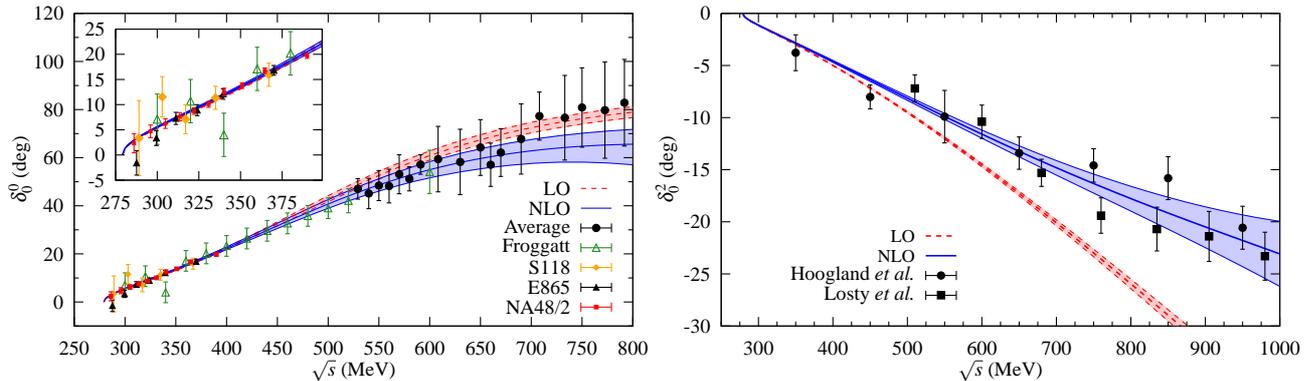
\centering
\includegraphics[height=5.cm,keepaspectratio]{fits_0.mps}
\includegraphics[height=5.cm,keepaspectratio]{fits_1.mps}
\caption{Comparison of our scalar $\pi\pi$ phase shifts to experimental data for $I=0$ (left panel) and $I=2$ (right panel). The (red) dashed  line shows our fit for the LO case ($V(s)\equiv V_2(s)$), whereas the (blue) solid  one shows the NLO fit ($V(s)=V_2(s)+V_4(s)$). The bands represent our uncertainties. The inset in the left panel shows in more detail the low energy $K_{e4}$ decays data. The data for $I=0$ are from the $K_{e4}$ decay data of Refs.~\cite{Rosselet:1976pu,Pislak:2001bf,Pislak:2003sv,Batley:2007zz,Batley:2010zza} (with isospin breaking effects taken into account as in \cite{Colangelo:2008sm}) and other data from Refs.~\cite{Froggatt:1977hu,ochsthesis,Grayer:1972,Estabrooks:1973,Kaminski:1996da,Hyams:1973zf,Protopopescu:1973sh}. For  $I=2$ the phase shifts are from Refs.~\cite{Losty:1973et,Hoogland:1977kt}.\label{fig:fitph}}
\end{figure*}
\begin{figure*}
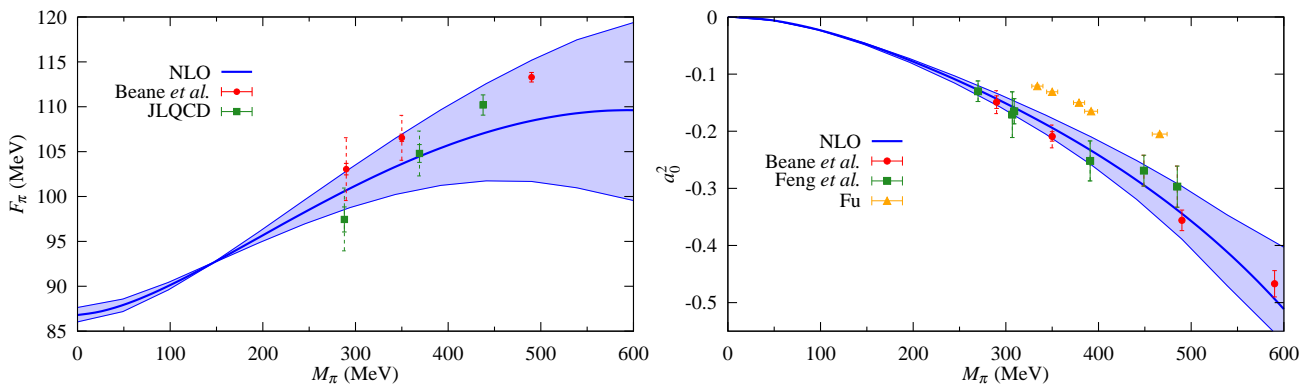
\centering
\includegraphics[height=5cm,keepaspectratio]{fits_8.mps}
\includegraphics[height=5cm,keepaspectratio]{fits_9.mps}
\caption{Dependence of $F_\pi$ (left panel) and $a_0^2$ (right panel) with $M_\pi$ as compared with lattice  QCD data. The (blue) solid  line is given by our NLO fit, whereas the band represents our estimated error. The data are from Refs.~\cite{Noaki:2008gx,Beane:2007xs,Feng:2009ij}. For $a_0^2$ we also show the data of Ref.~\cite{Fu:2011bz}, although we do not include them in our fits.
\label{fig:fitla}}
\end{figure*}

\begin{table}\centering
\caption{Summary of our LO and NLO fits. In the last column the $\chi^2$ per degree of freedom is given. 
\label{tab:fits_pars}}
\begin{tabular}{ccccccc}\hline\hline
Fit    & $a$     &$\bar{l}_1$ & $\bar{l}_2$ & $\vphantom{\frac{\frac{x}{y}}{\frac{x}{y}}}\bar{l}_3$ & $\bar{l}_4$ & $\displaystyle\frac{\chi^2}{\text{d.o.f.}}$\\ \hline
LO     & $-1.36 \pm 0.12$ & - & - & - & - & $1.6$\\
NLO    & $-1.2 \pm 0.4$ & $ 0.8 \pm 0.9$ & $4.6 \pm 0.4$ & $2 \pm 4$ & $3.9 \pm 0.5$ & $0.7$ \\ \hline\hline
\end{tabular}
\end{table}

At LO, there is just one free parameter corresponding to the subtraction constant in $G(s)$. At NLO, there are, in addition, four LECs, $\bar{l}_i,\ i=1,2,3,4$. For $I=0$, the phase shifts that we fit contain the very precise data of $K_{e4}$ decays below $\sqrt{s} = 400\ \textrm{MeV}$ \cite{Rosselet:1976pu,Pislak:2001bf,Pislak:2003sv,Batley:2007zz,Batley:2010zza}. These data are corrected for isospin breaking effects, as explained in Ref.~\cite{Colangelo:2008sm}. Above that energy, the data of Ref.~\cite{Froggatt:1977hu} and the average of different experiments \cite{ochsthesis,Grayer:1972,Estabrooks:1973,Kaminski:1996da,Hyams:1973zf,Protopopescu:1973sh}, as used \textit{e.g.} in Ref.~\cite{nd}, are taken into account. For $I=2$, the data come from Refs.~\cite{Losty:1973et,Hoogland:1977kt}. The fits extended to a maximum energy $\sqrt{s_\text{max}} = 0.8\ \text{GeV}$ at LO, both for $I=0$ and $I=2$, whereas at NLO we extend this range up to $\sqrt{s_\text{max}} = 1\ \text{GeV}$ for $I=2$. This is not done for $I=0$ because of the related presence of the $K\bar{K}$ threshold and 
the $f_0(980)$ resonance. The phase shifts are denoted by $\delta_0^I$, with $I=0,~2$. For our NLO fits we also fit recent lattice QCD results  as functions of the pion mass for $F_\pi$ \cite{Noaki:2008gx,Beane:2007xs} and the isotensor scalar scattering length, $a_0^2$ \cite{Beane:2007xs,Feng:2009ij}.\footnote{We consider the spread of these lattice QCD results as a source of systematic error for our fits. The final errors included in the fit are depicted by the dashed error bars in Fig.~\ref{fig:fitla}.} The dependence of $F_\pi$ with the pion mass is calculated at NLO in ChPT, Eq.~\eqref{eq:fpiNLO}. The scattering length $a_0^2$ is defined through the threshold expansion in powers of $\vp^2$ of our full results:
 \begin{equation}
\frac{\text{Re}T_{0}^{I}}{16\pi} = a_0^I + b_0^I \vp^2 + \mathcal{O}(|\vp|^4)~\text{,}
\end{equation} 
that we extrapolate in terms of the pion mass squared.

The resulting values for the fitted parameters are given in Table~\ref{tab:fits_pars}. At LO the subtraction constant for the $G(s)$ function is $a=-1.36 \pm 0.12$. Four LECs appear  additionally to the subtraction constant as free parameters at NLO. In order to avoid large correlation among them, the subtraction constant at NLO is constrained to remain near its value at LO. This is done by adding a new term to the $\chi^2$ taking into account the difference between the values of $a$ at LO and NLO, but  enlarging its error at LO from $0.12$ to  $0.2$, so that its contribution to the resulting $\chi^2$ is tiny but enough to remove the large correlations that would appear otherwise among the LECs and the subtraction constant. The parameters of both fits (LO and NLO) are shown in Table~\ref{tab:fits_pars}, and the corresponding phase shifts are plotted in Fig.~\ref{fig:fitph} with their respective errors. The left panel is for $I=0$ and the right one for $I=2$. The (red) dashed  lines arise from our fit at LO  ($V(s)\equiv V_2(s)$), whereas the (blue) solid  ones show the NLO fit ($V(s)=V_2(s)+V_4(s)$). In the inset of the upper panel  the agreement of our results with the lower energy data from $K_{e4}$ decay can be appreciated. We must stress that the difference between LO and NLO manifests mostly in the $I=2$ channel phase shifts, as can be seen in Fig.~\ref{fig:fitph}. In this channel, the left-hand cut is more important, but our amplitudes only incorporates the latter in a perturbative way, so that at NLO it is well reproduced, but it is absent at LO. In Fig.~\ref{fig:fitla} our results for $F_\pi$ (left panel) and $a_0^2$ (right panel) are shown, and compared with the aforementioned lattice QCD results.

\newcommand{\nohay}{\multicolumn{1}{c}{-}}
\begin{table*}\centering
\caption{Comparison of different phenomenological and lattice QCD determinations of the LECs $\bar{l}_i$, $i=1,2,3,4$. Together with every reference, for an easier comparison the initials of the authors or those of the collaboration are given.\label{tab:compLECS}}
\begin{tabular}{rlllll} \hline\hline
\multicolumn{2}{c}{Ref.} &\multicolumn{1}{c}{$\vphantom{\frac{{a^2}^2}{b}}\bar{l}_1$} & \multicolumn{1}{c}{$\bar{l}_2$} & \multicolumn{1}{c}{$\bar{l}_3$} & \multicolumn{1}{c}{$\bar{l}_4$} \\ \hline
\cite{Gasser:1983yg}          & GL          & $-2.3 \pm 3.7$ & $\hphantom{+} 6.0  \pm 1.3 $ & $\hphantom{+}2.9  \pm 2.4 $ & $\hphantom{+} 4.6 \pm 0.9$ \\
\cite{cola7}                  & CGL         & $-0.4 \pm 0.6$ & $\hphantom{+} 4.31 \pm 0.11$ & \nohay & $\hphantom{+} 4.4 \pm 0.2$ \\
\cite{ABT}                    & ABT         & $\hphantom{+}0.4 \pm 2.4$ & $\hphantom{+} 4.9 \pm 1.0$ & $\hphantom{+} 2.5^{+1.9}_{-2.4}$ & $\hphantom{+} 4.20 \pm 0.18$ \\
\cite{PePo}                   & PP          & $-0.3\pm 1.1$ & $\hphantom{+} 4.5 \pm 0.5$ & \nohay & \nohay \\
\cite{GKMS}                   & GKMS        & $\hphantom{+}0.37 \pm 0.95 \pm 1.71$ & $\hphantom{+} 4.17 \pm 0.19 \pm 0.43$ & \nohay & \nohay \\
\cite{BCT}                    & BCT         & \nohay & \nohay & \nohay & $\hphantom{+} 4.4 \pm 0.3$ \\
\cite{sroca}                  & OR          & \nohay & \nohay & \nohay & $\hphantom{+} 4.5 \pm 0.3$ \\
\cite{DFGS}                   & DFGS        & \nohay & \nohay & $-15 \pm 16$ & $\hphantom{+} 4.2 \pm 1.0$ \\ \hline
\cite{Batley:2010zza}         & NA48/2      & \nohay & \nohay & $\hphantom{+} 2.6 \pm 3.2$ & \nohay \\ \hline
\cite{RBCUKQCD}               & RBC/UKQCD   & \nohay & \nohay & $\hphantom{+} 2.57 \pm 0.18$  & $\hphantom{+} 3.83 \pm 0.9$ \\
\cite{PACSCS}                 & PACS-CS     & \nohay & \nohay & $\hphantom{+} 3.14 \pm 0.23$ & $\hphantom{+} 4.04 \pm 0.19$ \\
\cite{ETM}                    & ETM         & \nohay & \nohay & $\hphantom{+} 3.70 \pm 0.07 \pm 0.26$ & $\hphantom{+} 4.67 \pm 0.03 \pm 0.1$ \\
\cite{JLQCDTWQCD,JLQCDTWQCD2} & JLQCD/TWQCD & \nohay & \nohay & $\hphantom{+} 3.38 \pm 0.40 \pm 0.24^{+0.31}_{-0}$ & $\hphantom{+} 4.09 \pm 0.50 \pm 0.52$ \\
\cite{MILC}                   & MILC        & \nohay & \nohay & $\hphantom{+} 2.85 \pm 0.81^{+0.37}_{-0.92}$  & $\hphantom{+} 3.98 \pm 0.32^{+0.51}_{-0.28}$ \\ \hline
\multicolumn{2}{c}{This work}   & $\hphantom{+}0.8 \pm 0.9$ & $\hphantom{+} 4.6 \pm 0.4$ & $\hphantom{+} 2 \pm 4$ & $\hphantom{+} 3.9 \pm 0.5$ \\ \hline\hline
\end{tabular}
\end{table*}
\begin{figure*}
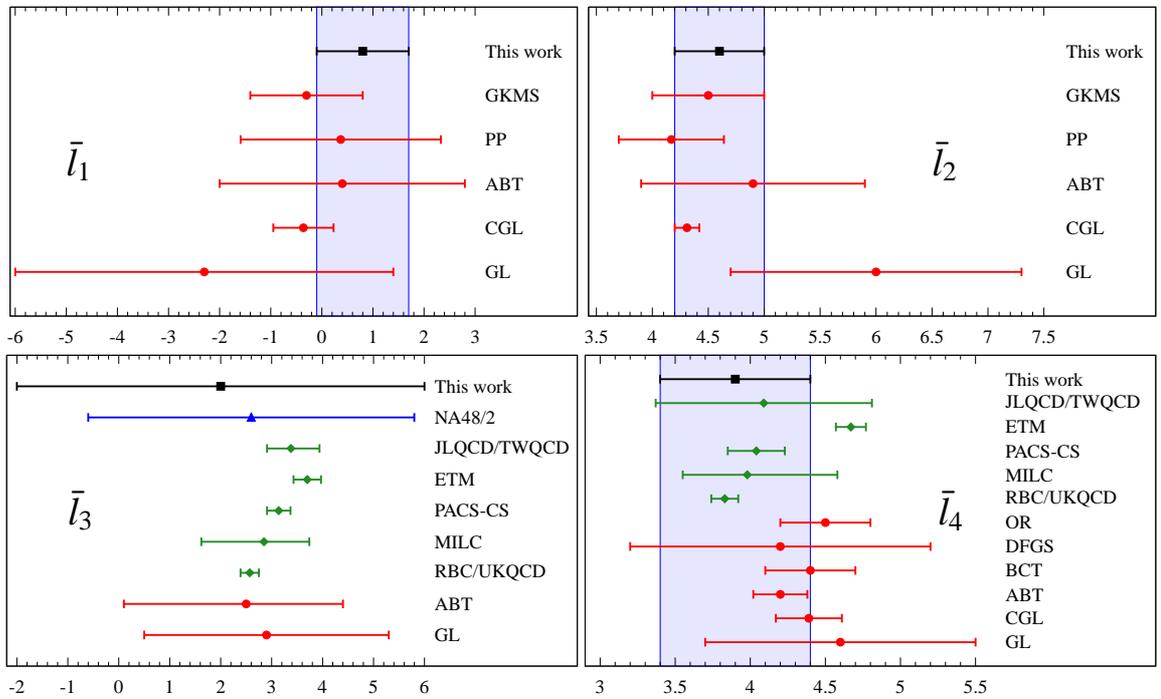
\centering
\includegraphics[width=0.42\textwidth,keepaspectratio]{lecs_3.mps}
\includegraphics[width=0.42\textwidth,keepaspectratio]{lecs_4.mps}\vspace{1mm}
\includegraphics[width=0.42\textwidth,keepaspectratio]{lecs_1.mps}
\includegraphics[width=0.42\textwidth,keepaspectratio]{lecs_2.mps}
\caption{Comparison of the different lattice QCD  and phenomenological  determinations of the LECs collected in Table~\ref{tab:compLECS}. The  (green) diamonds are lattice QCD determinations, and (red) circles  are the phenomenological ones. The range obtained for $\bar{l}_3$ by the NA48/2 Collaboration is represented by 
a  (blue) triangle. The (black) squares  are our results. For an easier comparison, we have included a shaded area that represents our results (except for $\bar{l}_3$).
\label{fig:compLECS}}
\end{figure*}

In Table~\ref{tab:compLECS} we collect some phenomenological \cite{Gasser:1983yg,cola7,PePo,BCT,ABT,GKMS,DFGS,sroca} and lattice QCD \cite{RBCUKQCD,JLQCDTWQCD,JLQCDTWQCD2,MILC,ETM,PACSCS} determinations of the LECs. For the latter the last values of each collaboration are taken, and, in addition, the direct $SU(2)$ fit results are selected if values for $SU(2)$ and $SU(3)$ fits are offered. 
 We have also included the range obtained for $\bar{l}_3$ from the data of the NA48/2 Collaboration \cite{Batley:2010zza}. These determinations are compared graphically in Fig.~\ref{fig:compLECS}, where for every LEC the different results are compatible within errors.
The lattice QCD results concerning $\bar{l}_{1,2}$ are scarce. The JLQCD and TWQCD Collaborations \cite{JLQCDTWQCD} recently reported $\bar{l}_1 - \bar{l}_2 = -2.9 \pm 0.9 \pm 1.3$, whereas, from our fit, we obtain $\bar{l}_1 - \bar{l}_2 = -3.8 \pm 1.3$. For the phenomenological determinations in Table~\ref{tab:compLECS}, since $\bar{l}_{1,2}$ agree  well between each other, also the aforementioned difference between these LECs does. We finally note that from our fit we obtain at NLO ChPT that $F=86.8 \pm 0.8\ \text{MeV}$, so that $F_\pi/F = 1.065 \pm 0.010$, compatible  with the estimate of lattice QCD results given in Ref.~\cite{reviewcola},  $F_\pi/F = 1.073 \pm 0.015$.

Our function $G(s)$ stems from the calculation of a once-subtracted dispersion relation (see \textit{e.g.} Ref.~\cite{nd}). If, instead, it is calculated approximately by employing a three-momentum cut-off $\Lambda$, one has the following relation between the subtraction constant and $\Lambda$ \cite{prdlong,plb}:
\begin{equation}\label{eq:a_cutoff}
a(\mu) = -1 + \log \frac{e \mu^2}{4 \Lambda^2} + \mathcal{O}\left(\frac{M_\pi^2}{\Lambda^2}\right)~\text{.}
\end{equation}
Our values for the fitted subtraction constant gives a cut-off $\Lambda \simeq 750\ \text{MeV}\simeq M_\rho$, which is quite a natural value. We will make use of these considerations based on Eq.~\eqref{eq:a_cutoff} later on, when dealing with the $M_\pi$ dependence of the $\sigma$ pole position.

\begin{table}[h]
\caption{$\sigma$ pole position and threshold parameters for the isoscalar scalar partial-wave amplitude.\label{tab:fits_pred}}
\begin{tabular}{rccccc}\hline\hline
Fit & $\sqrt{s_\sigma}\ \textrm{(MeV)}$                & & $a_0^0$           & & $b_0^0 M_\pi^2$ \\ \hline
 LO  & $465 \pm \hphantom{0}2 - i~231\pm\hphantom{0}7$ & & $0.209 \pm 0.002$ & & $0.278 \pm 0.005$ \\
NLO  & $440 \pm            10 - i~238\pm           10$ & & $0.219 \pm 0.005$ & & $0.281 \pm 0.006$ \\ \hline\hline
\end{tabular}
\end{table}

The $\sigma$ pole appears in the second  or unphysical Riemann sheet of the amplitude. This sheet is reached by changing the function  $G(s)$ in the following manner \cite{npa}. For $s$ real and above threshold we have
\begin{equation}\label{eq:gii}
G_\text{II}(s+i\epsilon) = G_\text{I}(s+i\epsilon) - \Delta G(s)~\text{,}
\end{equation}
where the subscript denotes the physical (I) or the unphysical (II) Riemann sheet. In the previous equation, $\Delta G(s)$ is the discontinuity along the unitarity cut,
\begin{equation}\label{eq:gii_2}
\Delta G(s) = G_\text{I}(s+i\epsilon) - G_\text{I}(s-i\epsilon) = -i\frac{p(s)}{8\pi\sqrt{s}}~\text{,}
\end{equation}
with $p(s)=\sqrt{\vp^2} = \sqrt{s/4-M_\pi^2}$, the CM pion three-momentum,  such that $\text{Im}\,p(s) > 0$. In order to explore the unphysical Riemann sheet, one then makes the analytical extrapolation in the cut complex $s$ plane of Eq.~\eqref{eq:gii}. 

In the second sheet the $\sigma$ resonance is a pole in the $I=0$ $S$-wave $\pi\pi$ amplitude,
\begin{equation}\label{eq:t_sigma_pole}
T_\text{II}(s\simeq s_\sigma) = -\frac{g_\sigma^2}{s-s_\sigma} + \cdots~\text{,}
\end{equation}
being $g_\sigma$ the coupling to the $\pi\pi$ channel and the ellipsis indicate the rest of terms in the Laurent series around $s_\sigma$ (with Im$s_\sigma <0$). The pole position $s_\sigma$ is given in Table~\ref{tab:fits_pred}, together with the resulting values for the threshold parameters of the scalar-isoscalar partial wave. The $\sigma$ pole position is used to define its mass and width, $M_\sigma -i\, \Gamma_\sigma/2\equiv \sqrt{s_\sigma}$. 

\begin{figure}
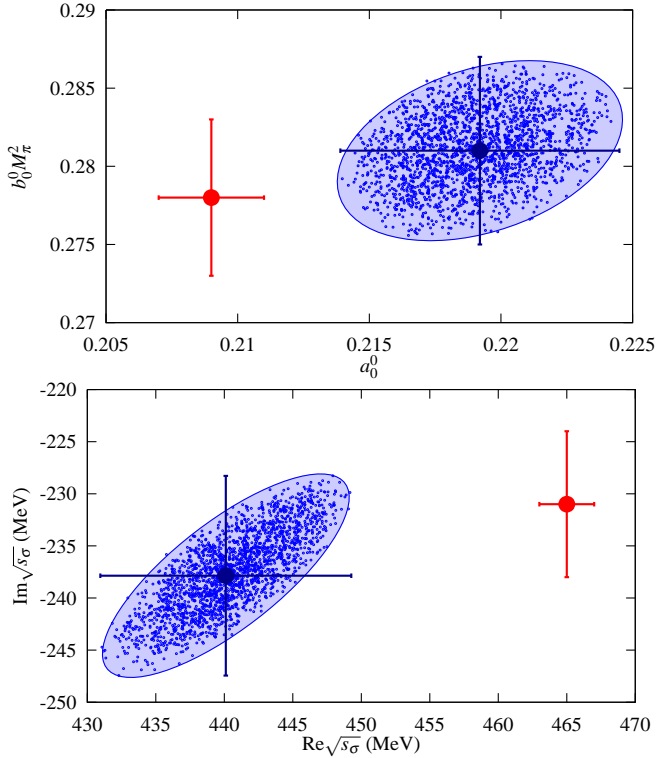
\centering
\includegraphics[height=5cm,keepaspectratio]{fits_2.mps}
\includegraphics[height=5cm,keepaspectratio]{fits_3.mps}
\caption{Montecarlo-like error analysis for the $\sigma$ mass ($M_\sigma \equiv \text{Re}\sqrt{s_\sigma}$) and half-width ($\Gamma_\sigma/2 \equiv -\text{Im}\sqrt{s_\sigma}$) and threshold parameters $a_0^0$ and $b_0^0$. The (blue) error ellipses correspond to the NLO fit while the single (red) point with errors is for the LO result. 
\label{fig:errors}}
\end{figure}

The error analysis for any quantities calculated here (\textit{e.g.} the fitted values for the LECs, $\sigma$ pole position, etc) is performed by randomly varying our parameters around their fitted values and accepting those values for the parameters which have a $\chi^2 < \chi^2_\text{min} + \Delta\chi^2$. Here $\chi^2_\text{min}$ is the best value for the $\chi^2$. For the LO case, since there is just one free parameter, we give our two-sigma confidence interval (otherwise the errors would be too small), given by $\Delta\chi^2=4$. At NLO the one-sigma confidence interval corresponds to $\Delta\chi^2=5.9$.  The resulting error ellipses are shown in Fig.~\ref{fig:errors} for the threshold parameters, upper panel, and for the $\sigma$ mass and width, lower panel. Notice that since there is only one free parameter at LO then a curve results instead of an error ellipse as in NLO. This is why at LO we have just shown the resulting value with its errors.


\subsection{The $\sigma$ meson. Comparison with other determinations}
\begin{table}
\caption{Values of $M_\sigma$, $\Gamma_\sigma/2$, $a_0^0$ and $b_0^0$ extracted from the literature. The value of Ref.~\cite{Batley:2010zza} corresponds to the latest experiment on $K_{e4}$ decays (with the errors added in quadrature for an easier comparison).
\label{tab:compare}}
\begin{tabular}{lllcc} \hline\hline
Ref. & $M_\sigma\ \text{(MeV)}$ & $\Gamma_\sigma/2\ \text{(MeV)}$ & $a_0^0$ & $b_0^0 M_\pi^2$ \\ \hline
\cite{zhou1}     & $470 \pm 50$ & $285 \pm 25$ & - & - \\
\cite{caprini}   & $441^{+16}_{-8}$ & $272^{+9}_{-13}$ & - & - \\
\cite{martin}    & $484 \pm 17$ & $255 \pm 10$ & $0.233 \pm 0.013$ & $ 0.285 \pm 0.012$\\
\cite{Albaladejo}& $456 \pm 12$ & $241 \pm 14$ & - & - \\
\cite{capri4}    & $463\pm 6^{+31}_{-17}$ & $254\pm6^{+33}_{-34}$ & $0.218\pm 0.014$ & $0.276\pm 0.013$\\
\cite{narison22}     & $452 \pm 12$           & $260\pm 15$ & - & - \\
\cite{kami8}     & $457^{+14}_{-13}$ & $279^{+11}_{-7}$ & - & - \\
\cite{Batley:2010zza}    &      \nohay     & \nohay    &  $0.222 \pm  0.014$ & - \\
\cite{leutyplb} & \nohay & \nohay & $0.220 \pm 0.005$ & $ 0.276 \pm 0.006$ \\ \hline
This work        & $440 \pm 10$ & $238 \pm 10$ & $0.219 \pm 0.005$ & $ 0.281 \pm 0.006$\\ \hline
Average          & $453 \pm \hphantom{0}5$ & $258 \pm \hphantom{0}5$ & $0.220 \pm 0.003$ & $ 0.279 \pm 0.003$ \\
Mean          & $458 \pm 14$ & $261 \pm 17$ & $0.223 \pm 0.007$ & $ 0.280 \pm 0.004$ \\ \hline\hline
\end{tabular}
\end{table}
\begin{figure*}
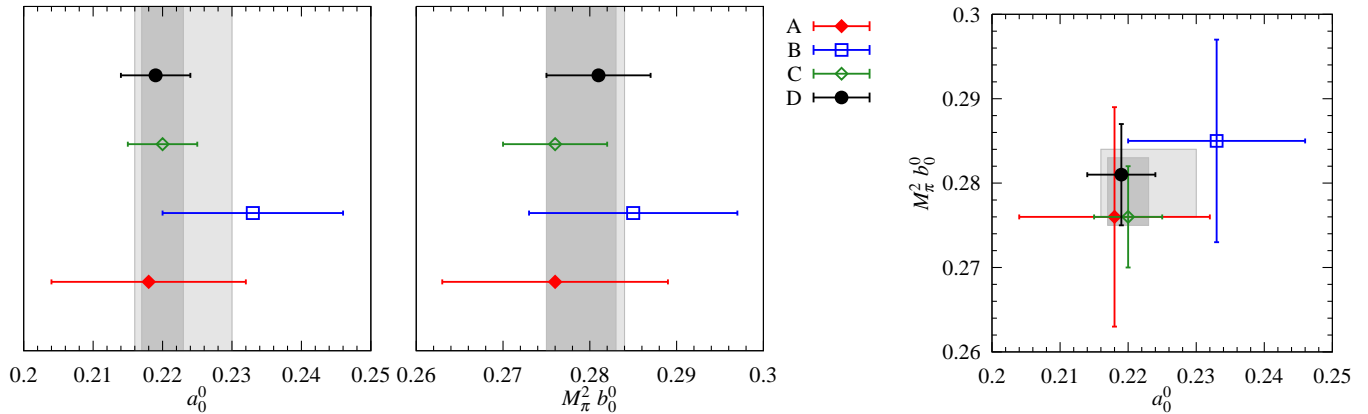
\centering
\includegraphics[height=5.4cm,keepaspectratio]{comp_sig_thr_5.mps}
\includegraphics[height=5.4cm,keepaspectratio]{comp_sig_thr_6.mps}
\includegraphics[height=5.4cm,keepaspectratio]{comp_sig_thr_4.mps}
\caption{In this figure we show the  values for the threshold parameters $a_0^0$ and $b_0^0$ from different papers in the literature, as indicated in the plots. In the first two panels, from left to right, the  (dark gray) inner strip corresponds to the interval covered by the weighted average whereas the (light gray) outer strip  is for the mean value, both given in Table~\ref{tab:compare}. In the last panel, the rectangles correspond to the aforementioned intervals  in the $a_0^0$--$b_0^0 M_\pi^2$ plane. The references are: A \cite{capri4}, B \cite{martin}, C \cite{leutyplb} and D refers to the NLO determination of this work.
\label{fig:comp_a_b}}
\end{figure*}
\begin{figure*}
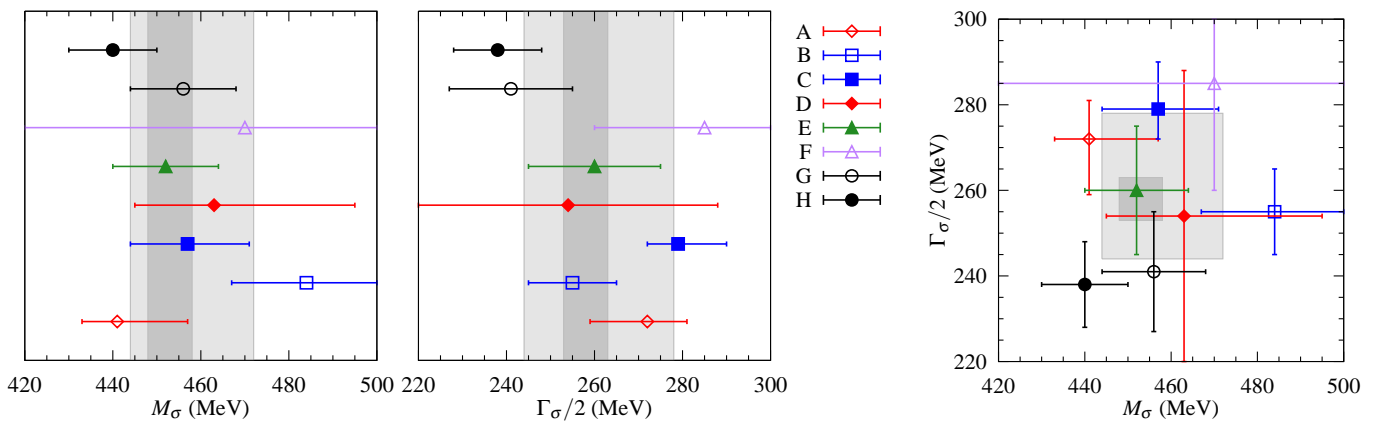
\centering
\includegraphics[height=5.45cm,keepaspectratio]{comp_sig_thr_2.mps}
\includegraphics[height=5.45cm,keepaspectratio]{comp_sig_thr_3.mps}
\includegraphics[height=5.45cm,keepaspectratio]{comp_sig_thr_1.mps}
\caption{In this figure we show the values for the mass and width of the $\sigma$ resonance  from different papers in the literature, as indicated in the plots. In the first two panels, from left to right, the (dark gray) inner strip  corresponds to the interval covered by the weighted average whereas the (light gray) outer strip  is for the mean value, both given in Table~\ref{tab:compare}. In the last panel, the rectangles correspond to the aforementioned intervals in the $M_\sigma$--$\Gamma_\sigma/2$ plane. The references are: A \cite{caprini}, B \cite{martin}, C \cite{kami8}, D \cite{capri4}, E \cite{narison22}, F \cite{zhou1}, G \cite{Albaladejo}, and F refers to the NLO determination of this work.
\label{fig:comp_sig}}
\end{figure*}

We compare now our results for the $\sigma$ mass and width as well as for the threshold parameters with other determinations from Refs.~\cite{zhou1,caprini,martin,Albaladejo,capri4,narison22,kami8}. References \cite{caprini,martin} are recent sophisticated determinations of the pion pole position claiming to be very precise.  In Ref.~\cite{Albaladejo}, based on chiral Lagrangians and the implementation of the N/D method, a detailed study of meson-meson scattering in the scalar sector up to around $\sqrt{s} = 2\ \text{GeV}$ was performed. All the relevant channels were taken into account, even the $4\pi$ channel through the $\sigma\sigma$ and $\rho\rho$ channels whose interactions kernels were predicted making use of chiral symmetry and vector meson dominance. A good description of the data considered was achieved, which allowed a full description of the resonances experimentally seen up to that energy.\footnote{In Table~\ref{tab:compare} we double the errors of our previous determination \cite{Albaladejo}, so as they have a similar size as those from other calculations. In this way the weighted average is not so much biassed from just one determination.}

The relevant quantities contained in those references are collected in Table~\ref{tab:compare}, and compared in Figs.~\ref{fig:comp_a_b} and \ref{fig:comp_sig} with our LO and NLO determinations. If all these determinations can be considered as different \textit{measures} of the same physical quantity, then they should be \textit{compatible}. A good check of their mutual compatibility is to determine whether they are compatible within errors with their weighted average.\footnote{For a given set of $N$ independent measures ${x_i}$ with their errors $\sigma_i$, the (weighted) average is given by $\bar{x} = \left(\sum_{i=1}^N x_i/\sigma_i^2\right)/\left(\sum_{i=1}^N 1/\sigma_i^2\right)$ and the standard deviation $\sigma$ by $1/\sigma^2 = \sum_{i=1}^N 1/\sigma_i^2$.} These values are calculated and given in Table~\ref{tab:compare}.

The ideal situation is that for the threshold parameters $a_0$ and $b_0$, as can be seen by simple inspection of Fig.~\ref{fig:comp_a_b}, or directly  from the values in Table~\ref{tab:compare}. All values agree within errors with their weighted average:
\begin{align}
a_0^0 & = 0.220 \pm 0.003~\text{,} \nn \\
b_0^0 M_\pi^2 & = 0.279 \pm 0.003~\text{.}
\label{thpvalues}
\end{align}
The latest NA48/2 Collaboration result \cite{Batley:2010zza} is $a_0^0 = 0.2220 \pm 0.0128_\text{stat} \pm 0.0050_\text{syst} \pm 0.0037_\text{th}$,  in good agreement with Eq.~\eqref{thpvalues}. For completeness we also report our result at NLO for the $I=2$ isoscalar scattering length:
\begin{align}
a_0^2&=-0.0424\pm 0.0012~\text{.}
\end{align}
The last value from $K_{e4}$ decays of the NA48/2 Collaboration \cite{Batley:2010zza} is $a_0^2=-0.0432\pm 0.0086_{\rm{stat}}\pm 0.0034_{\rm syst} \pm 0.0028_{\rm th}$, whereas the precise determination of Ref.~\cite{leutyplb} gives $a_0^2 = -0.0444 \pm 0.0010$. At this point, it is worth stressing that our unitarized amplitudes with the kernels calculated at NLO allow a good reproduction of the low energy behavior ($K_{e4}$ data and scattering lengths) while keeping the agreement with the higher energy data.

The case of the $\sigma$ mass and width is not so mild. In Fig.~\ref{fig:comp_sig} one can  see that the agreement within errors of the different values with the weighted average starts at the level of $(2-3)\sigma$. At this stage it is then preferable to take the mean of the different measures instead of the weighted average. In this way we have:
\begin{align}
M_\sigma & = 458 \pm 14\ \text{MeV}~\text{,}\nn\\
\Gamma_\sigma/2 & = 261 \pm 17\ \text{MeV}~\text{.}
\label{meansigma}
\end{align}
The resulting error is around 3 times bigger than that for the weighted average over the different values considered. The different determinations agree within errors with the above result, Eq.~\eqref{meansigma}. It can be concluded that our present knowledge on the pole position of the $\sigma$ meson is quite precise, with the uncertainty of the order of few tens of MeV, lying in a range much narrower than the values nowadays reported in the PDG.


\subsection{Dependence with $M_\pi$ of the $\sigma$ meson mass and width}
\label{sigma.mpi}\nocite{Albaladejo:2011zz}

\begin{figure}
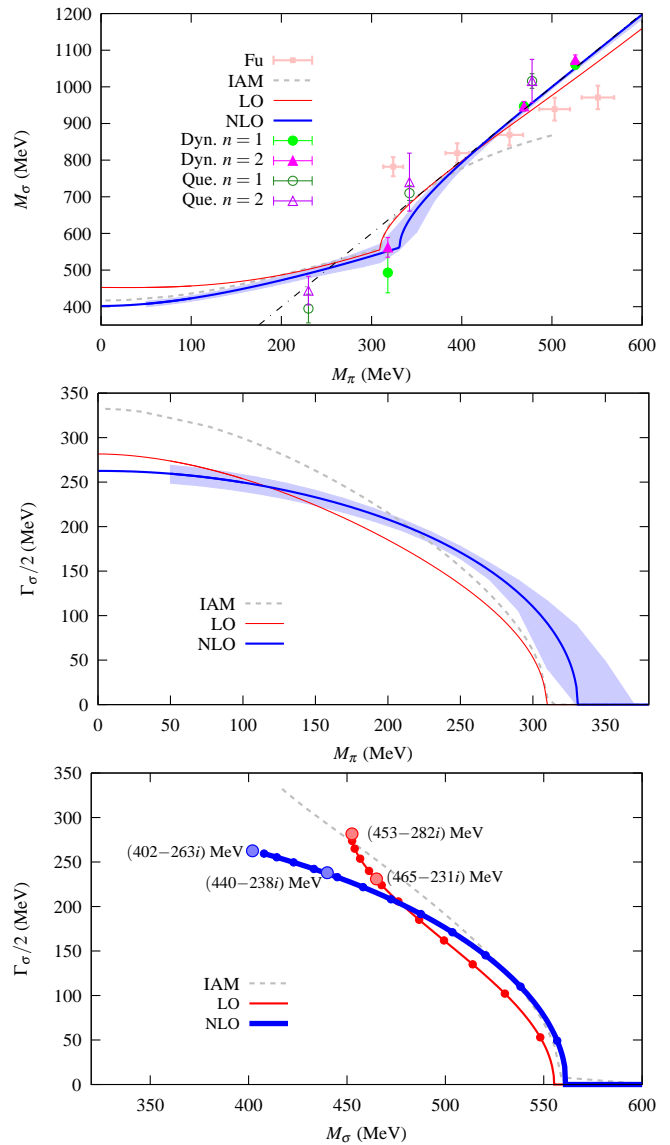
\centering
\includegraphics[height=5cm,keepaspectratio]{fit_chiral_limit_0.mps}
\includegraphics[height=5cm,keepaspectratio]{fit_chiral_limit_1.mps}
\includegraphics[height=5cm,keepaspectratio]{fit_chiral_limit_2.mps}
\caption{From top to bottom. First (second) panel: Mass (half width) of the $\sigma$ as a function of $M_\pi$. In the last panel we show the half-width as a function of the mass of the $\sigma$ while varying $M_\pi$. In the figures the (red) thinner  and (blue) thicker solid lines correspond to the LO and
NLO results, respectively. In the upper panel the (black) thin dot-dashed line represents the two-pion threshold, $2M_\pi$.  The larger circles in the last panel highlight the chiral limit and physical case results, whereas the smaller circles represent $25\ \text{MeV}$ steps in $M_\pi$, starting at $M_\pi=50\ \text{MeV}$. The dashed, gray lines are the results of Ref.~\cite{Pelaez:2010fj}. The squares in the first panel correspond to the lattice QCD 
results of Ref.~\cite{Fu:2011zzh}, while the rest of points are taken from  Ref.~\cite{Prelovsek:2010kg}.
\label{fig:ch_lim}}
\end{figure}

\begin{figure*}
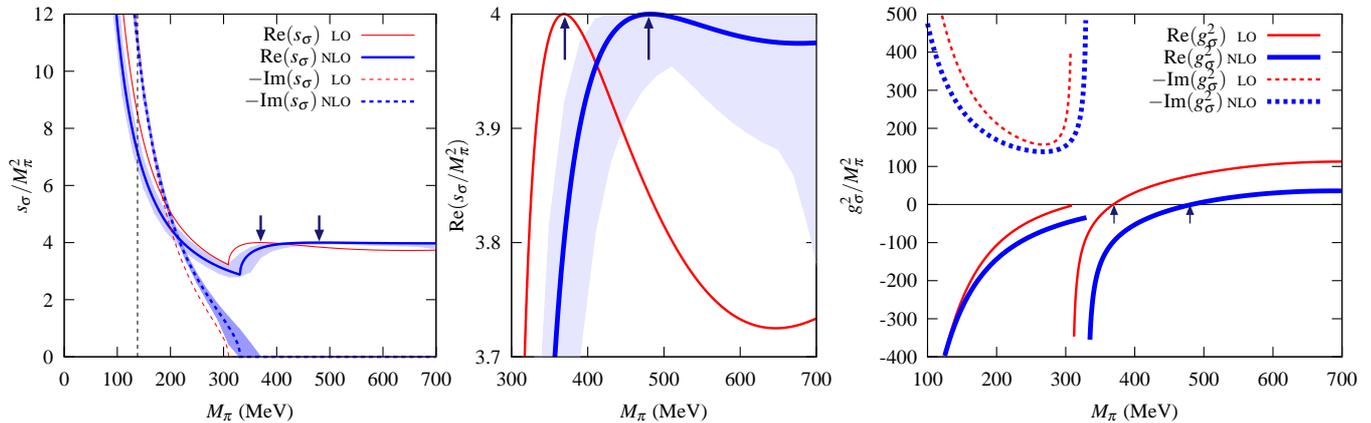
\centering
\includegraphics[height=5.5cm,keepaspectratio]{evolucion_sigma_1.mps}
\includegraphics[height=5.5cm,keepaspectratio]{evolucion_sigma_2.mps}
\caption{From left to right. In the first two panel we show $s_\sigma$ in units of the pion mass squared as a function of $M_\pi$. The second panel shows in more detail the region for $M_\pi\geq 300$~MeV. In the last panel $g_\sigma^2$ is depicted in the same units. In the figure the (red) thinner solid lines are for the LO results, and the (blue) thicker solid ones correspond to NLO. The solid lines correspond to the real part of the quantity shown, whereas the dashed ones represent its imaginary part. We indicate with arrows the points at which $s_\sigma=4M_\pi^2$ at LO and NLO.
\label{fig:evol_m_sig}}
\end{figure*}

We can now study the evolution of the $\sigma$ meson properties when the physical pion mass $M_\pi$ varies (\textit{e.g.} by changing the current quark masses in QCD). This is an interesting problem by itself. It is also related to the form factor of the $\sigma$ meson, $F_\sigma(s)$, since  $\text{d}s_\sigma / \text{d} M_\pi^2$ and $F_\sigma(0)$ are proportional by the Feynman-Hellman theorem, as discussed below. At LO, the only changes produced by varying $M_\pi^2$ are those occurring inside the kernel $V_2(s)$ and the loop function $G(s)$. At NLO, $F_\pi$ varies with $M_\pi^2$ because of Eq.~\eqref{eq:fpiNLO}, and also the LECs because it follows from Eq.~\eqref{eq:lecsNLO} that:
\begin{equation}
\bar{l}_i\left(M_\pi^2\vphantom{M_{\pi\text{, phys}}^2}\right) = \bar{l}_i\left(M_{\pi\text{, phys}}^2\right) - \log \frac{M_\pi^2}{M_{\pi\text{, phys}}^2}~\text{.}
\end{equation}
We can consider the subtraction constant $a$ in the function $G(s)$ as independent of $M_\pi$ in view of Eq.~\eqref{eq:a_cutoff}. With the above considerations one searches  the $\sigma$ pole position in the $s$-complex plane, $s_\sigma$,  for different values of $M_\pi$, just as in the physical pion mass case. The coupling $g_\sigma^2$ is also obtained  by means of the Cauchy theorem.

Before discussing this evolution, it is useful to make some analytical derivations. Let us consider the unitarized $\pi\pi$ amplitude,  Eq.~\eqref{eq:tpipi_uni},  as a function of both the Mandelstam variable $s$ and the pion mass squared, $T(s,M_\pi^2)$. In the second Riemann sheet it reads:
\begin{equation}\label{eq:tsm2}
T(s,M_\pi^2) = \frac{V(s,M_\pi^2)}{1+V(s,M_\pi^2)G_{II}(s,M_\pi^2)}~\text{.}
\end{equation}
This function  has a Laurent series around $s_\sigma$ expressed in Eq.~\eqref{eq:t_sigma_pole}. 
Taking the derivative of $T(s,M_\pi^2)$ with respect to $M_\pi^2$ in both sides of Eq.~\eqref{eq:t_sigma_pole}, and attending to the double-pole terms, one obtains:
\begin{equation}\label{eq:dssig}
\dot{s}_\sigma(M_\pi^2) = -\frac{g_\sigma^2(M_\pi^2)}{V(s_\sigma,M_\pi^2)^2}\left( \dot{V}(s_\sigma,M_\pi^2) - V(s_\sigma,M_\pi^2)^2 
\dot{G}_{II}(s_\sigma,M_\pi^2)  \right)~\text{,}
\end{equation}
where the dot denotes derivative with respect to $M_\pi^2$. In the previous equation we have taken into account that Eq.~\eqref{eq:tsm2} requires  that $G_{II}(s_\sigma)=-1/V(s_\sigma)$ at the pole position $s_\sigma$.

Analogously, since $g_\sigma(M_\pi^2)^2$ is minus the residue of the pole  of the amplitude in the $s$ variable, one gets:
\begin{equation}\label{eq:gmpi2}
g_\sigma^2(M_\pi^2) = \frac{V(s_\sigma,M_\pi^2)^2}{V'(s_\sigma,M_\pi^2)-V(s_\sigma,M_\pi^2)^2G'_{II}(s_\sigma,M_\pi^2)}~\text{,}
\end{equation}
where the prime denotes a derivative with respect to the $s$ variable. One should replace $G_{II}(s_\sigma,M_\pi^2)$ by $G(s_\sigma,M_\pi^2)$ (the function in the physical Riemann sheet)  in Eqs.~\eqref{eq:dssig} and \eqref{eq:gmpi2} for the case when the $\sigma$ pole becomes a bound state. From Eqs.~\eqref{eq:dssig} and \eqref{eq:gmpi2}, given the knowledge of $s_\sigma$ and $g_\sigma^2$ in the physical case, the evolution of the pion pole and the coupling with $M_\pi^2$ could be studied directly. We have checked that the numerical results are the same as those obtained by looking for the pole in the complex plane for different pion masses, as explained above.

The main features of the evolution of the $\sigma$ meson with $M_\pi$ can be grasped by the inspection of Figs.~\ref{fig:ch_lim} and  \ref{fig:evol_m_sig}.   In Fig.~\ref{fig:ch_lim}  we show $\sqrt{s_\sigma}$  as a function of $M_\pi$, so that, $M_\sigma$ is shown is the upper plane, $\Gamma_\sigma/2$ in the middle one and the plane $M_\sigma$--$\Gamma_\sigma/2$ in the panel on the bottom. The (red) thinner solid lines originate from the LO calculation, $V=V_2$, and the (blue)  thicker solid ones from the NLO results, $V=V_2+V_4$, Eq.~\eqref{eq:v4}. For the physical situation ($M_\pi \simeq 140\ \text{MeV}$), we have the case just described, that is, the $\sigma$ meson is seen as a pole in the unphysical Riemann sheet. As we increase $M_\pi$, the imaginary part of $\sqrt{s_\sigma}$ decreases, becoming zero at $M_\pi \simeq 310\ \text{MeV}$ for LO and at $M_\pi \simeq 330\ \text{MeV}$ for NLO.\footnote{At this point  another pole (not shown in the figures) starts to appear below the $\sigma$ one. This is due to the appearance of two real solutions for the equation $1 + V(s)G(s) = 0$, since the imaginary part of $s_\sigma$ is zero in this region. There is no need to consider further this pole since, irrespectively of whether it lies in the same Riemann sheet than the higher pole, the effects of the latter overwhelmingly dominate over those of the former. For smaller $M_\pi$, since the solutions are not real, the $\sigma$ corresponds to two  complex conjugated values.}

In Fig.~\ref{fig:evol_m_sig} we show $s_\sigma$ in units of the pion mass squared in the first and second panels from left to right. In the latter the scale of the  ordinate axis changes and is restricted to values slightly slower than $4 M_\pi^2$, so that one can appreciate the evolution of the real part of $s_\sigma$ and distinguish it from the line $s_\sigma=4\,M_\pi^2$ (which is difficult to realize from the first panel for $M_\pi\geq 300$~MeV). In the last panel we show $g_\sigma^2$ in the same units for varying $M_\pi$. For all the panels the solid (dashed) lines are for the real (imaginary) part, and the thicker (thinner) lines correspond to NLO (LO) results. Notice that both for LO and NLO, $g_\sigma^2$ diverges at the point where $s_\sigma$ becomes purely real. Approaching this point from lower values of $M_\pi$, $\text{Im}\ g_\sigma^2$ diverges, whereas, approaching it from higher values of $M_\pi$ then $\text{Re}\ g_\sigma^2$ is the one that diverges. This can be understood from the behavior of the derivative of $s_\sigma$, that is not defined precisely at this point, and in view of Eq.~\eqref{eq:dssig}, where it is seen that $\dot{s}_\sigma \propto g_\sigma^2$.

For even larger values of $M_\pi$ ($M_\pi \simeq 370\ \text{MeV}$ at LO and $M_\pi \simeq 480\ \text{MeV}$ at NLO), $s_\sigma$ osculates the $2\pi$ threshold, while standing below it, and changes from the unphysical Riemann sheet to the physical one, becoming a bound state. Since $s_\sigma \simeq 4M_\pi^2$ close to this point, the binding energy is small, and so is the coupling, becoming exactly zero when $s_\sigma = 4M_\pi^2$.  These points are indicated with arrows in Fig.~\ref{fig:evol_m_sig}.  This behavior can be shown analytically. From Eq.~\eqref{eq:gmpi2}, one deduces that for $s_\sigma \simeq 4M_\pi^2$,
\begin{equation}
g_\sigma^2 = -\eta\ 64\pi M_\pi \sqrt{\left\lvert s_\sigma - 4M_\pi^2 \right\rvert}~\text{,}
\label{couplingthreshold}
\end{equation}
with $\eta = +1$ for the unphysical Riemann sheet (at the left of this point) and $\eta = -1$ for the physical Riemann sheet (at the right). Therefore, $g_\sigma^2=0$ for $s_\sigma = 4M_\pi^2$, as indicated by the arrows in the rightmost panel of Fig.~\ref{fig:evol_m_sig}. However, it is worth noticing that from Eq.~\eqref{couplingthreshold} it follows that $g^2_\sigma/\sqrt{|s_\sigma/4-M_\pi^2|}\equiv g^2_\sigma/|\mathbf{p}_\sigma|$ is finite. On the other hand, the fact that the pole changes from one Riemann sheet to the other  in a continuous way can be understood in terms of Eqs.~\eqref{eq:gii} and \eqref{eq:gii_2}. The difference between the $G(s)$ function calculated in the two Riemann sheets is given by a piece  proportional to $\sigma(s_\sigma) = \sqrt{1-4M_\pi^2/s_\sigma}$ that vanishes for $s_\sigma = 4M_\pi^2$. At this point, where the $\sigma$ is a zero bound state, one also has an infinite value for the scattering length. 

\begin{figure}\centering
\includegraphics[height=5cm,keepaspectratio]{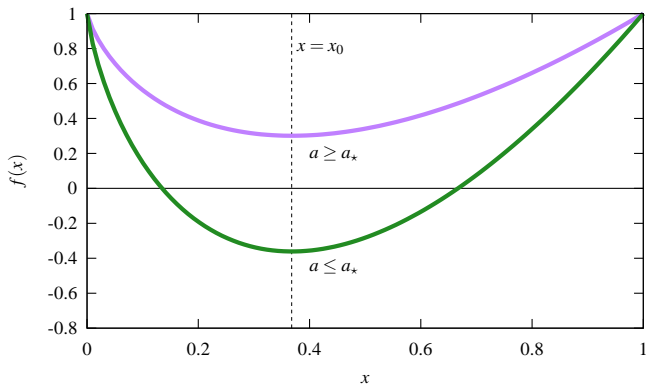}
\caption{Representation of the function $f(x)$, Eq.~\eqref{eq:fcrit}, for two values of $a$, $a > a_\star$ (upper line) and $a < a_\star$  (bottom line). The variable $x$ is defined in Eq.~\eqref{eq:xcrit}.
\label{fig:fcrit}}
\end{figure}

The mere existence of this \textit{critical} point can be examined analytically. For $s=4M_\pi^2$, the function $G(s)$ can be written as:
\begin{equation}
G(s=4M_\pi^2) = \frac{a + \log \frac{M_\pi^2}{\mu^2}}{16\pi^2} \equiv \frac{\log\frac{M_\pi^2}{\mu_a^2}}{16\pi^2}~\text{,}
\end{equation}
with $\mu_a^2 = e^{-a}\mu^2$ a new scale. If we concentrate on the simpler case of LO, $V(4M_\pi^2) =  7M_\pi^2/2 F_\pi^2$, the equation for finding a pole at $s=4M_\pi^2$, $V^{-1} + G = 0$, can be cast as $f(x)=0$, with
\begin{equation}~\label{eq:fcrit}
f(x) = 1 + \alpha x \log x~\text{,}
\end{equation}
where  
\begin{align}
x= M_\pi^2/\mu_a^2
\label{eq:xcrit}
\end{align}
 and 
\begin{equation}\label{eq:alphadef}
\alpha = 7\mu_a^2/(32\pi^2 F_\pi^2) \geqslant 0~.
\end{equation}
Since  $\alpha \geqslant 0$ a zero of the $f(x)$ function is only possible for $0 \leqslant x \leqslant 1$. Actually two zeros of this function exists  if the value of the function at its minimum $x_0 = e^{-1}$ is negative (see Fig.~\ref{fig:fcrit}). This condition in terms of the variable $a$ requires that 
the latter is smaller than the critical value $a_\star$,  
\begin{equation}
a_\star = -1 + \log \frac{7 \mu^2}{32\pi^2F_\pi^2}~\text{.}
\end{equation}
If this is the case there is a zero for $0<x<x_0$ and another one for $x_0 < x < 1$. For our value of the renormalization scale, $\mu=770\ \text{MeV}$,  $a_\star \simeq -0.6$, so that the fitted value $a \simeq -1.4$ given in Table \ref{tab:fits_pars} is much smaller than $a_\star$. We also have that our value for $x_0$ corresponds to $M_\pi \simeq 900$~MeV, then a pole with $s_\sigma= 4M_\pi^2$ exists for $0 < M_\pi < 900\ \text{MeV}$. The solution of Eq.~\eqref{eq:fcrit} for the value of $a$ fitted gives that this pole is located at $M_\pi \simeq 370\ \text{MeV}$, as stated above and indicated by the left most arrow in the panels of Fig.~\ref{fig:evol_m_sig}.

For the NLO case, the situation becomes somewhat more involved, and the function $f(x)$ is now:
\begin{equation}\label{eq:fcritNLO}
f(x) = 1 + \alpha(x) x \log x \Big( 1 + \alpha(x) \beta(x) x \Big)~\text{,}
\end{equation}
where $\alpha(x)$ is defined as in  Eq.~\eqref{eq:alphadef}, but at NLO one has to take into account its implicit dependence  
 on $x \propto M_\pi^2$ through $F_\pi$. On the other hand, $\beta(x)$ is defined as
\begin{align}
\beta(x) & =  \frac{40}{147}L - \frac{2}{7}\log x~\text{,} \nn \\
L & = \bar{l}_1^p + 2\bar{l}_2^p - \frac{3}{8}\bar{l}_3^p + \frac{21}{10}\bar{l}_4^p + \frac{21}{8} + \frac{189}{40}\log x_p~\text{,}
\end{align}
where $\bar{l}_i^p\equiv \bar{l}_i(M^2_{\pi,{\rm phys}})$ corresponds to the LECs calculated at the physical pion mass and 
\begin{align}
x_p=M_{\pi,{\rm phys}}^2 /\mu_a^2~.
\end{align}
 For the values collected in Table~\ref{tab:fits_pars} we find that $s_\sigma = 4M_\pi^2$ for $M_\pi \simeq 480\ \text{MeV}$. Nevertheless,  this value 
is quite sensitive to the LECs, and it should be taken merely as indicative (for some values of the LECs not far from the 
  fitted ones the change from virtual to bound state does not occur at all). This sensitivity is illustrated by the error band in Figs.~\ref{fig:ch_lim} and \ref{fig:evol_m_sig}.

In Fig.~\ref{fig:ch_lim} our results  on the pion mass dependence of the $\sigma$ pole position, partially presented in Ref.~\cite{Albaladejo:2011zz}, are compared with other works. The (gray) dashed line, denoted by IAM, gives the results of Ref.~\cite{Pelaez:2010fj} in the framework of the IAM. The points shown come from the lattice QCD studies of Refs.~\cite{Prelovsek:2010kg,Fu:2011zzh}. Interestingly, we find a remarkably good agreement with the curve from the IAM results \cite{Pelaez:2010fj} for $M_\pi \lesssim 400$~MeV. As stated by the authors, the point where $s_\sigma = 4M_\pi^2$, and thus the $\sigma$ meson becomes a bound state, is $M_\pi \simeq 460\ \text{MeV}$ when they employed the NLO ChPT amplitudes \cite{Pelaez:2008}, whereas $M_\pi \simeq 290$--$350\ \text{MeV}$ when the N$^2$LO ChPT amplitudes were used. We show in Fig.~\ref{fig:ch_lim} the curves of Ref.~\cite{Pelaez:2010fj} corresponding to this latter case. 

A lattice QCD search of light scalar tetraquarks with $J^{PC}=0^{++}$ (we focus here on the $I=0$ results) is performed in Ref.~\cite{Prelovsek:2010kg}. 
Along with the lowest $\pi(\vp)\pi(-\vp)$ scattering state, an additional lighter state is found. For the dynamical simulations of Ref.~\cite{Prelovsek:2010kg} the former state is denoted in Fig.~\ref{fig:ch_lim} with $n=1$ (green filled circles) and the latter one with $n=2$ (pink filled triangles). For the quenched simulations we use the (green) empty circles and the (purple) empty triangles, in the same order as before.\footnote{However, we must also point out that the lattice QCD simulations are performed for each pion mass at a single volume and lattice spacing, so the continuum and infinite volume values of the $\sigma$ meson mass in the bound state case may differ from those values.} The points with $n=1$ and 2 overlap at each pion mass, and the quantitative agreement with our curves is satisfactory. However, both our curves and the lattice QCD results of Ref.~\cite{Prelovsek:2010kg} do not agree with most of the points of the lattice QCD calculation of Ref.~\cite{Fu:2011zzh} and, in addition, the tendency of the points is qualitatively different to that for our results and those of Ref.~\cite{Prelovsek:2010kg}. 

For larger values of $M_\pi$ we obtain values for the $\sigma$ meson mass, both at LO and NLO,  that remain below but always close to the $\pi\pi$ threshold, in agreement with the lattice QCD results of Ref.~\cite{Prelovsek:2010kg}. Note that this is not the case for the IAM calculation of Ref.~\cite{Pelaez:2010fj} for $M_\pi\gtrsim 400$~MeV.  The fact that the $\sigma$ meson follows so closely the threshold for higher values of $M_\pi$, both according to our calculation and to the lattice QCD calculation of Ref.~\cite{Prelovsek:2010kg}, clearly indicates  that for such masses it is dynamically generated from the $\pi\pi$ interactions. We elaborate further on the nature of the $\sigma$ resonance below. However, one should keep in mind that the $\sigma$ meson becomes an anti-bound or virtual state between those pion masses 
in which it has zero width and has not crossed to the physical Riemann sheet yet.  In the bound state case, an additional state appears in the energy levels spectrum in the box, whereas an anti-bound state does not. In order to discern the latter situation one should look at other computable quantities, such as the sign of the $I=0$ $S$-wave $\pi\pi$ scattering length.

It is also interesting to study the chiral limit, $M_\pi \to 0$. As can be seen in Fig.~\ref{fig:evol_m_sig}, $s_\sigma/M_\pi^2 \to \infty$, because $s_\sigma$ remains finite in this limit. Indeed, the  values calculated for $s_\sigma$ near the chiral limit behave as  (for $M_\pi \leqslant 150$~MeV),
\begin{equation}
s_\sigma(M_\pi^2) = s_{\sigma,\chi} + a\, M_\pi^2 + b\, M_\pi^2 \log \frac{M_\pi^2}{M^2_{\pi,\text{phys}}}~\text{,}
\label{ch.limit}
\end{equation}
with the values of the $\sigma$ pole position in the chiral limit
given by  $\sqrt{s_{\sigma,\chi}} = 453 - i\, 282~\text{MeV}$ (LO) and $\sqrt{s_{\sigma,\chi}} = 402 - i\, 263\ \text{MeV}$ (NLO), 
see  Fig.~\ref{fig:ch_lim}.


\section{The scalar form factor of the $\sigma$ meson}\label{sec:sigmaff}

We turn now our attention to the calculation of the scalar form factor of the $\sigma$ meson, that is, the interaction of the $\sigma$ resonance with a scalar source (denoted in the following by $H$).  As an intermediate step we calculate first the scattering of two pions in the presence of a scalar source, from which we extract the scalar form factor of the $\sigma$. This can be done because the  $\sigma$ originates as a pole in the interaction of a scalar isoscalar pair of pions, as discussed in Sec.~\ref{sec:pipiscattering}. We start by considering in Sec.~\ref{subsec:kinematics} the kinematics of the $\pi\pi H \to \pi\pi$ reaction, which is  somewhat more complicated than the standard kinematics of a two-body reaction. In Sec.~\ref{subsec:pipistopipi}, we discuss the one-loop calculation of the amplitude $\pi\pi H \to \pi\pi$ from the chiral Lagrangians of Sec.~\ref{sec:chilag}. In terms of this  amplitude one can derive the scalar form factor of the $\sigma$ meson, as performed in  Sec.~\ref{subsec:formfactor}. This is accomplished by taking into account pion rescattering, similarly as done above  for $\pi\pi$ scattering, with some modifications that are carefully examined.

\subsection{Kinematics}
\label{subsec:kinematics}

\begin{figure}
\centering
\includegraphics[height=4.5cm,keepaspectratio]{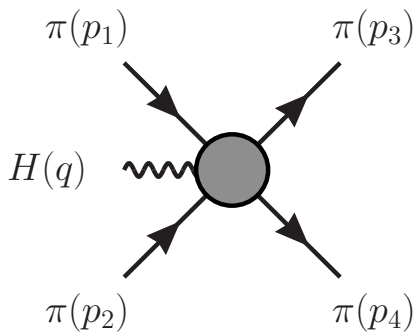}
\caption{Kinematics of the $\pi\pi$ scattering process in the presence of a scalar source, $\pi(p_1)\pi(p_2) H(q)\to \pi(p_3)\pi(p_4)$. Pions correspond to the solid lines and the scalar source to the wavy one. The gray blob indicates the interactions involved. 
\label{fig_kin}}
\end{figure}

\begin{figure}
\centering
\includegraphics[height=8cm,keepaspectratio]{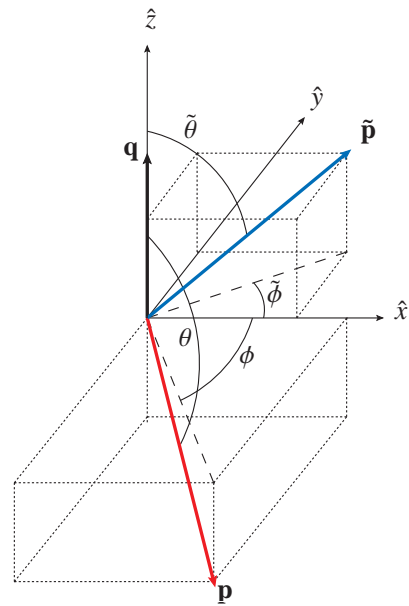}
\caption{The unit three-momenta in terms of the polar and azimuthal angles.
\label{fig:angles}}
\end{figure}

We are interested in pion-pion scattering with a scalar source, $ \pi(p_1) + \pi(p_2) + H(q)\to \pi(p_3) + \pi(p_4)$,  Fig.~\ref{fig_kin}. The overall center-of-mass frame, CM, is the same as the rest frame  of the final pions, while that corresponding to the initial ones is denoted by CMB. Due to the presence of the scalar source CMB does not coincide with CM. In the CM one has
\begin{align}
p_3 &= \left( \frac{\sqrt{s}}{2} , +\vp \right)\text{,}\nn\\
p_4 & = \left( \frac{\sqrt{s}}{2} , -\vp \right)\text{,}\nn\\
\vp^2 & = \frac{s}{4} - M_\pi^2~\text{,}
\end{align}
and
\begin{align}
\vp_1 + \vp_2 &= -\vq \nn\\
p_1^0 + p_2^0 &= \sqrt{s} - q^0\nn\\
q&\equiv (q^ 0,\vq)~.
\end{align}
We denote by $s$ and $s'$ the invariant masses squared for the final and initial pions, in order. At the end of the calculation, the limit $s,~s' \to s_\sigma$ is taken. It follows that 
\begin{equation}
(p_1+p_2)^2 = s' = \left( \sqrt{s} - q^0 \right)^2 - {\vq\,}^2 = s + q^2 - 2 q^0 \sqrt{s}~,
\end{equation}
and then,
\begin{align}
q^0 & = \frac{s-s'+q^2}{2\sqrt{s}}~,\nn\\
{\vq}^2& = \frac{(s+s'-q^2)^2}{4s} - s'~.
\label{cm.q0q2}
\end{align}

Analogously, one has in CMB:
\begin{align}
p_1 & = \left(\frac{\sqrt{s'}}{2},+\widetilde{\vp}\right)~\text{,}\nn\\
p_2 & = \left(\frac{\sqrt{s'}}{2},-\widetilde{\vp}\right)~\text{,}\nn\\
\widetilde{\vp}^2 & = \frac{s'}{4} - M_\pi^2~\text{,}\nn \\ 
\widetilde{q}^0   & = \frac{s-s'- q^2}{2\sqrt{s'}}~\text{,}\nn\\
\widetilde{\vq}^2 & = \frac{(s+s'-q^2)^2}{4s'}-s~\text{.}
\end{align}
In the following  quantities with a tilde are expressed in CMB. Notice that $\widetilde{\vp}$ is the three-momentum of the  first pion in CMB, while $\vp$ refers to the three-momentum of the third pion in CM.

The final (initial) two-pion states are projected into $S$-wave in CM (CMB) because the $\sigma$ resonance is defined as a pole in the second Riemann sheet of the $\pi\pi$ isoscalar $S$-wave in CM (CMB). The unit three-momenta (indicated with a hat) are given in terms of the polar and azimuthal angles (see Fig.~\ref{fig:angles}) as:
\begin{eqnarray}
\uni{\vp} & = & \left( \sin\theta\cos\phi, \sin\theta\sin\phi, 
\cos\theta \right)~, \nn\\
\uni{\widetilde{\vp}} & = & \left( \sin\tilde{\theta}\cos\tilde{\phi}, \sin\tilde{\theta}\sin\tilde{\phi}, \cos\tilde{\theta} \right)~,\nn \\
\uni{\vq} & = & (0,0,1)~,
\end{eqnarray}
where we have chosen the $z$-axis to be the direction pointed by ${\uni{\vq}}$. We now work out the  Lorentz transformation from CMB to CM:
\begin{eqnarray}
\left(p_1 + p_2 \right)_{CMB} & = & \left( \sqrt{s'}, \vec{0} \right)~,\nn \\
\left(p_1 + p_2 \right)_{CM} & = & \left( \sqrt{s}-q^0, -{\vq\,} \right)=(\frac{s+s'-q^2}{2\sqrt{s}}, -{\vq\,})~.
\end{eqnarray}
The transformation reads:
\begin{eqnarray}
\frac{s+s'-q^2}{2\sqrt{s}} & = & \gamma \sqrt{s'}~,\nn \\
-{\vq\,} & = &  -\gamma\sqrt{s'}\mathbf{v}~.
\end{eqnarray}
It follows then that $\gamma=1/\sqrt{1-\mathbf{v}^2}$ and $\mathbf{v}$ are 
\begin{eqnarray}
\gamma & = & \frac{s+s'-q^2}{2\sqrt{s}\sqrt{s'}}~, \\
\mathbf{v} & = & \frac{{\vq\,}}{\gamma \sqrt{s'}} = \frac{2  \sqrt{s}}{s+s'-q^2}\vq~.
\label{lor.tra}
\end{eqnarray}
We further define the four-momenta $\Sigma$ and $\Delta$ given by
\begin{eqnarray}
\label{def.sigma.delta}
\Sigma & \equiv &\left( p_1 + p_2 \right) ~, \nn \\
\Delta & \equiv & \left( p_1 - p_2 \right)~. 
\end{eqnarray}
In the CM 
\begin{align}
\Sigma = \left( \sqrt{s} - q^0,-{\vq\,} \right)~.
\label{sigma.cm}
\end{align}
 The momentum transfer $\Delta$ has a simple expression 
in CMB where it is given by $\Delta=(0,2\widetilde{\vp})$. We then perform its Lorentz transformation to CM, with the result
\begin{eqnarray}
\Delta^0 & = &  - 2\frac{\vq \cdot \widetilde{\vp}}{\sqrt{s'}}~, \nn\\
\mathbf{\Delta} & = & 2\widetilde{\vp} + 2\left(\gamma-1\right) \abs{\widetilde{\vp}} \left( \uni{\vp} \cdot \uni{\vq} \right) \uni{\vq}~.
\label{delta.cm}
\end{eqnarray}

The problem has six independent Lorentz invariant kinematical variables.\footnote{One the five four-momenta involved in the reaction is fixed by energy-momentum conservation. From the other four ones we can construct 6 independent scalar products. Notice that $p_1^2=p_ 2^2=p_3^2=M_\pi^2$ and that $q^2$ can be derived from Eq.~\eqref{relation} in terms of other Lorentz invariants.} We define, in analogy with two body scattering, the following six alike Mandelstam variables,
\begin{align}
s &= (p_3 + p_4)^2 ~\textrm{,}\nn\\
s'& = (p_1 + p_2)^2~\textrm{,} \nn \\
t &= (p_1 - p_3)^2 ~\textrm{,}\nn\\
t' & = (p_2 - p_4)^2~\textrm{,}\nn  \\
u &= (p_1 - p_4)^2 ~\textrm{,}\nn\\
u'& = (p_2 - p_3)^2~\textrm{.}
\label{invariants}
\end{align}
 These variables fulfill the relationship
\begin{equation}
s + t + u + s' + t' + u' = q^2 + 8 M_\pi^2~,
\label{relation}
\end{equation}
which is the analogous one to $s + t + u = 4M_\pi^2$ valid for two-pion scattering, Eq.~\eqref{eq:stu}. Though $q^2$ and the variables in Eq.~\eqref{invariants} are not independent because of Eq.~\eqref{relation}, it is convenient to write the different amplitudes $\pi\pi H \to \pi\pi$  in terms of all of them, given the symmetries  present in the calculation.

In virtue of the previously worked Lorentz transformation, Eq.~\eqref{lor.tra}, we have the four-momenta properly defined in CM in terms of the key variables $s$, $s'$, $q^2$ and the polar and azimuthal angles in the two-pion center of mass frames (the Lorentz invariants only depend on the difference between the azimuthal angles, see Eq.~\eqref{eq:abc} below). It is convenient to express $p_1=(\Sigma+\Delta)/2$ and $p_2=(\Sigma-\Delta)/2$, with $\Sigma$ and $\Delta$ given in CM by Eqs.~\eqref{sigma.cm} and \eqref{delta.cm}. In terms of this set of variables, the Lorentz invariants of Eq.~\eqref{invariants} are given by
\begin{eqnarray}
t\phpr & = &  2M_{\pi}^2 - 2\left(\alpha + A + B + C  \right)~,\nn \\
t'     & = &  2M_{\pi}^2 - 2\left(\alpha - A - B + C  \right)~,\nn \\
u\phpr & = &  2M_{\pi}^2 - 2\left(\alpha + A - B - C  \right)~,\nn \\
u'     & = &  2M_{\pi}^2 - 2\left(\alpha - A + B - C  \right)~,
\label{exp.inv}
\end{eqnarray}
where
\begin{align}
\alpha &=\frac{1}{2} \Sigma^0  \cdot p_{3,4}^0=\frac{1}{8}(s+s'-q^2)~, \nn\\
A      &= \frac{1}{2}\Delta^0     \cdot p_{3,4}^0 = -\frac{1}{2}\abs{{\vq\,}}\abs{\widetilde{\vp}} \frac{\sqrt{s}}{\sqrt{s'}}  
\cos\tilde{\theta}~, \nn\\
B      &=-\frac{1}{2} \vec{\Sigma} \cdot \vp = +\frac{1}{2}\abs{{\vq\,}}\abs{\vp} \cos\theta~, \label{eq:abc} \\
C      &=-\frac{1}{2} \vec{\Delta}\cdot \vp
=-|\vp||\widetilde{\vp}|\left(\uni{\vp}\cdot\uni{\widetilde{\vp}}+ \frac{(\sqrt{s}-\sqrt{s'})^2-q^2}{2\sqrt{s}\sqrt{s'}}\cos\theta \cos\tilde{\theta}\right)~. \nn
\end{align}

In the previous equation the five kinematical variables, $s$, $s'$, $q^2$, $\cos \theta$, $\cos\tilde{\theta}$ are used together with  the scalar product 
\begin{align}
\hat{\vp}\cdot \uni{\widetilde{\vp}}= \sin\theta\,\sin\tilde{\theta}\cos(\phi-\tilde{\phi})+\cos\theta\,\cos\tilde{\theta}~.
\end{align}

In terms of the  variables in Eq.~\eqref{invariants} one can  express the inverses of several pion propagators that appear in many  Feynman diagrams that contain the scalar source attached to an external pion leg, {\it cf.}  diagram (a.2) of Fig.~\ref{fig:pipis}. It results:
\begin{align}
D_1&=(q+p_1)^2 - M_{\pi}^2  =   s\phpr + t'     + u'     - 4M_\pi^ 2~\text{,} \nn\\
D_2&=(q+p_2)^2 - M_{\pi}^2  =   s\phpr + t\phpr + u\phpr - 4M_\pi^ 2~\text{,} \nn\\
D_3&=(q-p_3)^2 - M_{\pi}^2  =   s'     + t'     + u\phpr - 4M_\pi^ 2~\text{,} \nn\\
D_4&=(q-p_4)^2 - M_{\pi}^2  =   s'     + t\phpr + u'     - 4M_\pi^ 2~\text{.}
\label{eq:ds.def}
\end{align}
Because of Eqs.~\eqref{eq:abc}, the angular dependence of these inverse propagators is rather simple: $D_1$ and $D_2$ depend just on $\cos\tilde{\theta}$, while $D_3$ and $D_4$ do  on $\cos\theta$. The propagating pion can become on-shell for certain angles, giving rise to a pole in the propagators. These poles, when the $S$-wave angular projections are performed, result in logarithmic divergences. In particular, there is always a pole for $q^2 \to 0$. We treat this issue later on.

\subsection{The $\pi\pi \,H \to \pi\pi$ scattering amplitude}
\label{subsec:pipistopipi}

\begin{figure}
\centering
\includegraphics[height=9cm,keepaspectratio]{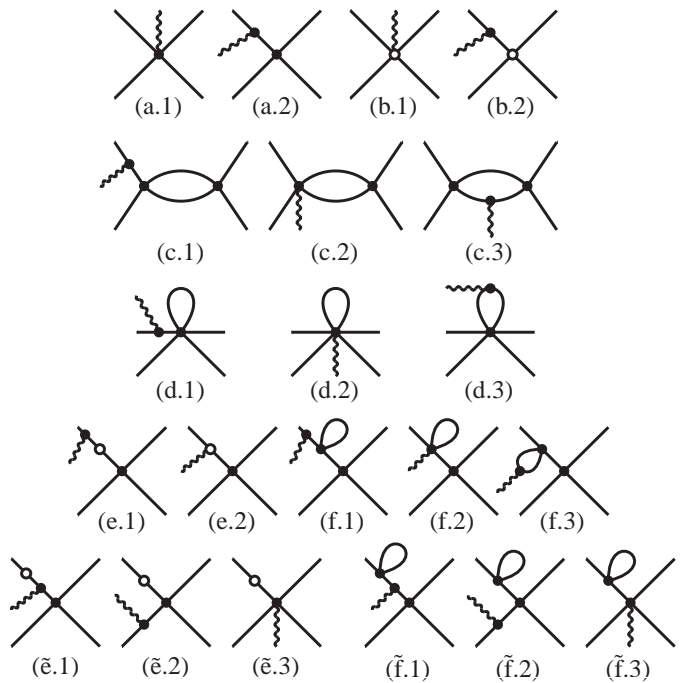}
\caption{\small Feynman diagrams for the $\pi\pi$ scattering amplitude in the presence of a scalar source, $\pi\pi\,H\to\pi\pi$, at one-loop order in ChPT.
\label{fig:pipis}}
\end{figure}

\begin{figure}
\centering
\includegraphics[width=0.45\textwidth,keepaspectratio]{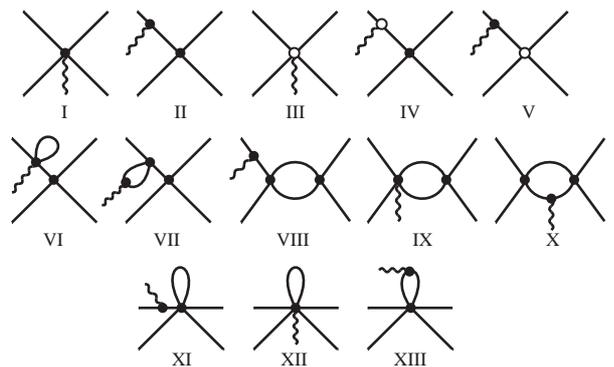}
\caption{Final set of Feynman diagrams for the $\pi\pi$ scattering in the presence of a scalar source, $\pi\pi\,H\to\pi\pi$, at ${\cal O}(p^4)$ in ChPT omitting the pion propagator dressing and wave function renormalization of the leading order diagrams in Fig.~\ref{fig:pipis}.\label{fig:pipis_def}}
\end{figure}

To determine the Feynman diagrams required for the $\pi\pi$ scattering  in the presence of a scalar source up to ${\cal O}(p^4)$ in ChPT it is useful to have in mind first those diagrams of plain  $\pi\pi$ scattering in Sec.~\ref{sec:pipiscattering}, Fig.~\ref{fig:pipi}. Now, one external scalar source has to be added in all the possible ways to those diagrams. As deduced from the Lagrangians ${\cal L}_2$ and ${\cal L}_4$, Eq.~\eqref{eq:lagGL}, the scalar source can couple to  any even number of pions.  In Fig.~\ref{fig:pipis} we show the diagrams that must be calculated at the one-loop level, where the external scalar source is indicated by a wiggly line. The LO diagrams correspond to  (a.1) and (a.2).\footnote{Of course, the scalar 
source can be attached to any of the pion legs but for conciseness we draw explicitly the attachment to only one. This should be understood in the following.} 
Diagrams (a.2), (e.1) and (f.1) can be handled together because their sum correspond 
to taking the full pion propagator in between the external source and the four-pion vertex, Eqs.~\eqref{pp.ren} and \eqref{mpi2}. In addition, all the diagrams on the bottom line of Fig.~\ref{fig:pipis}, namely, (\~{e}.1)--(\~{f}.3), correspond to the wave function renormalization of the LO ones. Both issues are derived to NLO from the pion self-energy diagrams, Fig.~\ref{fig:pse}, Eq.~\eqref{mpi2}. Once the  the renormalization of the pion propagator and the the wave function renormalization are taken into account,  as well as the rest of diagrams diagrams in Fig.~\ref{fig:pipis}, one has the basic topologies shown in  Fig.~\ref{fig:pipis_def}.

Compared with $\pi\pi$ scattering the presence of the $c$-number external scalar source $H$ complicates considerably the simple expressions for the former \cite{Gasser:1983yg}. The calculation for each of the diagrams in Fig.~\ref{fig:pipis_def} is given in Appendix~\ref{app:amp}. Specifically, we calculate the processes $\pi^0(p_1)\pi^0(p_2) H(q)\to\pi^0(p_3)\pi^0(p_4)$ and $\pi^0(p_1)\pi^0(p_2) H(q)\to\pi^+(p_3)\pi^-(p_4)$, with the former denoted by $T_{nn}$ and the latter by $T_{nc}$. These two processes are considered in order to isolate the pion pairs with  definite isospin ($I$) by taking the appropriate linear combinations. The standard decomposition of the $\pi^0\pi^0$ and $\pi^+\pi^-$ states in two-pion isospin definite states, $|\pi\pi(I I_3)\ra$, being $I_3$ the third-component of isospin, is 
\begin{align}
|\pi^0\pi^0\ra&=\sqrt{\frac{2}{3}} |\pi\pi(2 0)\ra  -  \sqrt{\frac{1}{3}} |\pi\pi(00)\ra~,\nn\\
|\pi^+\pi^-\ra&=-\sqrt{\frac{1}{6}} |\pi\pi(2 0)\ra  -  \sqrt{\frac{1}{2}} |\pi\pi(10)\ra  -  \sqrt{\frac{1}{3}} |\pi\pi(00)\ra ~,\nn
\end{align}
where we have taken into account that $|\pi^+\ra=-|\pi;I=1\, I_3=-1\ra$, as follows from the definition of the $\pi^+$ field, Eq.~\eqref{pif.def}. 
Because of isospin conservation (the scalar source $H(q)$ is isoscalar), the Wigner-Eckart theorem implies
\begin{align}
\la\pi^0\pi^0|{\cal S}|\pi^0\pi^0\,s\ra = & +\frac{2}{3}\la\pi\pi(2 0)|{\cal S}|\pi\pi(20)H\ra\nn\\
&+\frac{1}{3}\la\pi\pi(00)|{\cal S}|\pi\pi(00)H\ra~,\nn\\
\la\pi^+\pi^-|{\cal S}|\pi^0\pi^0\,s\ra = & -\frac{1}{3}\la\pi\pi(2 0)|{\cal S}|\pi\pi(20)H\ra\nn\\
&+\frac{1}{3}\la\pi\pi(00)|{\cal S}|\pi\pi(00)H\ra~, \label{eq:amplis_nn_nc}
\end{align}
with ${\cal S}$ the $S$-matrix. From this equation we can isolate the purely $I=0$ matrix element, ${\cal A}(s,s',q^2,\theta,\tilde{\theta},\phi,\tilde{\phi})$, corresponding to 
\begin{equation}
{\cal A}(s,s',q^2,\theta,\tilde{\theta},\phi,\tilde{\phi}) \equiv \la\pi\pi(0 0) \lvert {\cal S} \rvert \pi\pi(0 0)H\ra  ~.
\end{equation}
 From Eq.~\eqref{eq:amplis_nn_nc}, we have:
\begin{align}
{\cal A}(s,s',q^2,\theta,\tilde{\theta},\phi,\tilde{\phi}) & =\la\pi^0\pi^0|{\cal S}|\pi^0\pi^0\,s\ra+2\la\pi^+\pi^-|{\cal S}|\pi^0\pi^0\,s\ra \nn\\
& = T_{nn} + 2 T_{nc}~\text{.}\label{amp.def}
\end{align}
We are interested in this matrix element because  the $\sigma$ is isoscalar.

\begin{figure}
\centering
\includegraphics[width=.475\textwidth]{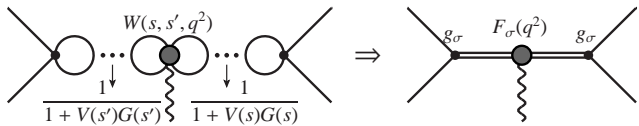}
\caption{External scalar source coupled to a double $\sigma$ pole in the $\pi\pi\, H\to\pi\pi$ process. The $\sigma$ pole is originated by the resummation of pion re-scattering, as indicated in the left diagram by the iteration of the unitarity two-point function.\label{fig:sss}}
\end{figure}

The $\sigma$ is an $S$-wave resonance so that it is also required the $S$-wave angular projection of the initial and final isoscalar pion pairs. This is straightforward for the final pions because the CM coincides with its own rest frame, with the result:
\begin{align}
|\pi\pi;00\ra_{34}&\equiv\frac{1}{4\pi}\int d\hat{\vp} \, |\pi(p_3)\pi(p_4)(00)\ra~.
\label{swave.34}
\end{align}
Regarding the initial pair of pions, its state is defined in CMB analogously as in the previous expression. One has still to perform the Lorentz boost to the CM frame so that
\begin{align}
|\pi\pi;00\ra_{12}&\equiv U(\vv)\frac{1}{4\pi}\int d\hat{\widetilde{\vp}} \, |\pi(\frac{\sqrt{s'}}{2},\widetilde{\vp})\pi(\frac{\sqrt{s'}}{2},-\widetilde{\vp})(00)\ra~,
\label{uv1}
\end{align}
where $U(\vv)$ is the Lorentz boost operator from  CMB to CM, with the velocity $\vv$ given in Eq.~\eqref{lor.tra}. When acting on the pion states (which have zero spin) the only effect is  the transformation of the four-momenta from CMB to CM. Then, 
we can also write Eq.~\eqref{uv1} as
\begin{align}
|\pi\pi;00\ra_{12}&=\frac{1}{4\pi}\int d\hat{\widetilde{\vp}} \, |\pi(p_1(\widetilde{\vp}))\pi(p_2(\widetilde{\vp}))(00)\ra~\text{,}
\label{swave.12}
\end{align}
where $p_1$ and $p_2$ are written in terms of the four-momenta in CMB. From Eq.~\eqref{def.sigma.delta}, $p_1=(\Sigma+\Delta)/2$, $p_2=(\Sigma-\Delta)/2$ with $\Sigma$ and $\Delta$  given in Eqs.~\eqref{sigma.cm} and \eqref{delta.cm}, in order, as a function of the CMB kinematical variables.

Employing the states projected in  $S$-wave, Eqs.~\eqref{swave.34} and \eqref{swave.12}, we are then ready to calculate  the required matrix element  in ChPT, $\varphi(s,s',q^2)$: 
\begin{equation}\label{eq:angularprojectionpipis}
\varphi(s,s',q^2)\equiv \frac{1}{32\pi^2}\int \text{d}^2 \Omega\ \mathcal{A}(s,s',q^2,\theta,\tilde{\theta},\phi,\tilde{\phi})~\text{.}
\end{equation}
Note that the extra factor $1/2$ in Eq.~\eqref{eq:angularprojectionpipis} arises because of the  unitary normalization, as explained after  Eq.~\eqref{eq:norm_uni}. In the last equation, the double solid angle integration is 
\begin{equation}
\int \text{d}^2 \Omega = \int_{-1}^{1}\!\!\!\text{d}\cos\theta\int_{-1}^{1}\!\!\!\text{d}\cos\tilde{\theta}\int_{0}^{2\pi}\!\!\!\text{d}\phi\int_{0}^{2\pi}\!\!\!\text{d}\tilde{\phi}~.
\label{ang.dob}
\end{equation}
One linear combination of azimuthal angles, $\phi$ and $\tilde{\phi}$, is a spare variable, and then one integration in Eq.~\eqref{ang.dob} is trivial. This is so because they appear just through the expression $\cos(\phi-\tilde{\phi})$, as explained above, Eq.~\eqref{eq:abc}. In fact, for any periodic angular function, $f(\gamma) = f(\gamma + 2\pi)$, one has:
\begin{equation}
\int_0^{2\pi}\!\!\!\!\!\!\mathrm{d}\phi \int_{0}^{2\pi}\!\!\!\!\!\!\mathrm{d} \tilde{\phi}\ f(\phi-\tilde{\phi}) 
= 2\pi \int_{0}^{2\pi}\!\!\! \!\!\!\mathrm{d} \gamma\ f(\gamma)~.
\label{triv.int}
\end{equation}

\subsection{Scalar form factor}
\label{subsec:formfactor}

Once the perturbative amplitude for the process $\pi\pi \,H \to \pi\pi$ is calculated,  we proceed by taking into account  pion rescattering, similarly as  was done for $\pi\pi \to \pi\pi$, see Eqs.~\eqref{eq:tpipi_uni}-\eqref{eq:v4}.  The resulting amplitude is denoted by $T_S(s,s',q^2)$, and following the same unitarization method as in Sec.~\ref{sec:pipiscattering} from Refs.~\cite{nd,plb}, it can be written as:
\begin{equation}\label{eq:resummation}
T_S(s,s',q^2) = \frac{W(s,s',q^2)}{\left(1+V(s)G(s) \right)\left(1+V(s')G(s') \right)}~\text{.}
\end{equation}

This is the analog to Eq.~\eqref{eq:tpipi_uni} but now for the process $\pi\pi\,H\to \pi\pi$, with the new kernel $W(s,s',q^2)$  instead of $V(s)$ in Eq.~\eqref{eq:tpipi_uni}.  It is important to  stress the presence of two factors $1+VG$ in the denominator of Eq.~\eqref{eq:resummation}. This is so because in  $\pi\pi\, H \to \pi\pi$ the presence of the scalar source $H(q)$ makes necessary to resum the unitarity loops corresponding to both final and initial state interactions. 

  The kernel $W(s,s',q^2)$ is obtained in a chiral expansion by matching Eq.~\eqref{eq:resummation} order by order with its perturbative calculation. 
  The chiral expansion of the kernel is 
\begin{align}
\label{wexp}
W = W_2 + W_4 + {\cal O}(p^6)~,
\end{align} 
where we omit the dependence on the arguments $s$, $s'$ and $q^2$ for easy reading. The subscripts in Eq.~\eqref{wexp} refer to the chiral order. Then, the amplitude Eq.\eqref{eq:resummation} is expanded,  as it was done in Eq.~\eqref{eq:chiralexpansion}, so that one has:
\begin{align}
T_S(s,s',q^2)  = & W_2 + W_4 \nn \\- & W_2V_2(s)G(s) - W_2V_2(s')G(s') + {\cal O}(p^6) \nn \\
 = & \varphi_2 + \varphi_4+{\cal O}(p^6)~\text{,}
\end{align}
where $\varphi_n(s,s',q^2)$ is the ${\cal O}(p^n)$ contribution to $\varphi(s,s',q^2)$ defined in Eq.~\eqref{eq:angularprojectionpipis}. The kernels $W_n(s,s',q^2)$ are determined by matching the above expressions order by order, so that:
\begin{align}\label{exp.wn}
W_2(s,s',q^2) & = \varphi_2(s,s',q^2) \nn\\
W_4(s,s',q^2) & = \varphi_4(s,s',q^2) + \varphi_2\,\xi_2(s)G(s) + \varphi_2\,\xi_2(s')G(s')~\text{,}
\end{align}
where it was used that $V_2(s) = \xi_2(s)$, Eq.~\eqref{eq:v4}. 

The form factor of the $\sigma$ meson, $F_\sigma(q^2)$, can now be extracted from $T_S(s,s',q^2)$, employing $W=W_2+W_4$ in Eq.~\eqref{eq:resummation}. For that one has to isolate the double $\sigma$ pole present in $T_S(s,s',q^2)$, as drawn on the right-hand side of Fig.~\ref{fig:sss}. The double $ \sigma$-pole contribution can be written as:\footnote{Because of invariance under temporal inversion the amplitudes for $\pi\pi\to \sigma$ and $\sigma\to\pi\pi$ are equal.}
\begin{equation}\label{eq:def_ff_dp}
 \frac{F_\sigma(q^2)\ g_\sigma^2}{(s-s_\sigma)(s'-s_\sigma)} = \lim_{s,s'\to s_\sigma} \frac{W(s,s',q^2)}{\left(1+V(s)G(s) \right)\left(1+V(s')G(s') \right)}
\end{equation}

Expanding the r.h.s. of the above equation around $s,s' \to s_\sigma$, and equating  the double-pole term, the result is:
\begin{equation}\label{eq:def_ff}
F_\sigma(q^2) = \frac{g_\sigma^2}{V(s_\sigma)^2} W(s_\sigma,s_\sigma,q^2)~\text{.}
\end{equation}

In determining the kernels $W_n(s,s',q^2)$, we have followed the master guidelines of pure $\pi\pi$ scattering procedure to take into account the rescattering of the pions given in Sec.~\ref{subsec:pipiampli}. However, some modifications are needed in our case because of the presence of the pion propagators in the external pion legs attached to a scalar source, see Fig.~\ref{fig:pipis_def}. Let us focus, for clearness, in the LO amplitudes $\varphi_2(s,s',q^2)$, corresponding to the diagrams I and II in Fig.~\ref{fig:pipis_def} (the amplitudes are given in Appendix~\ref{app:amp}). Before the angular projection in Eq.~\eqref{eq:angularprojectionpipis}, one has
\begin{equation}\label{eq:TSLO}
\mathcal{A}_2(s,s',q^2,\theta,\tilde{\theta},\phi,\tilde{\phi}) = -\frac{2B}{F_\pi^2}\left(1- 2\sum_{i=1}^{4}\frac{s_i-M_\pi^2/2}{D_i} \right)~\text{,}
\end{equation}
where the subscript in ${\cal A}$ refers to the chiral order, $s_{1,2} = s'$ and $s_{3,4} = s$, and the $D_i$ are the inverse of the pion propagators  given in Eq.~\eqref{eq:ds.def}. These contributions proportional to the propagators stem from the piece of diagram II in which the on-shell part of the $4\pi$ vertex is retained, so that the pion propagator is not cancelled out by an off-shell part from the $4\pi$ vertex ({\it cf.} Ref.~\cite{npa}). Considering, for conciseness, the case $s=s'$ (the one interesting for the $\sigma$ scalar form factor for which $s=s'=s_\sigma$), these propagators can be written as:
\begin{equation}
\frac{1}{D_1} = \frac{1}{\frac{q^2}{2} - 2\lvert \vp \rvert \lvert \vq \rvert \cos\tilde{\theta}}~\text{,}
\end{equation}
and similarly for the other $D_i$. It should be noted that for certain values of $q^2$ and $s$, these propagators can have a pole in the variable $\cos\tilde{\theta}$. In particular, for $q^2\to0$ this is always the case. Upon angular integration, this contribution gives rise to  an imaginary part that  diverges as $1/\sqrt{\lvert q^2 \rvert}$ for $q^2 \to 0^-$. As shown below, this limit is the one that matters  in order to calculate the quadratic scalar radius of the $\sigma$, but this divergence would lead to an undetermined value for it. This fact is not acceptable and  indicates a deficiency in the procedure followed up to now. 

\begin{figure}
\includegraphics[width=.475\textwidth]{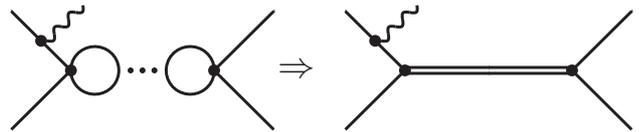}
\caption{String of unitary diagrams to be resummed when the scalar source is attached to an external leg. On the right, the resulting single $\sigma$ pole contribution is depicted.\label{fig:notformfactor}}
\end{figure}

Let us clarify this important technical point and the way it can be solved. The term of the amplitude ${\cal A}_2(s,s',q^2,\theta,\tilde{\theta},\phi,\tilde{\phi})$ in Eq.~\eqref{eq:TSLO} that is proportional to the pion propagators, that we denote by $\mathcal{A}_{2,\text{prop}}$,\footnote{Recall that we are interested in the $s=s'$ case.} can be written as:
\begin{equation}
\mathcal{A}_{2,\text{prop}} =   \xi_2(s)\sum_{i=1}^4 \frac{4B}{D_i}~\text{,}
\end{equation}
where we have taken into account that $\xi_2(s) = (s-M_\pi^2/2)/F_\pi^2$. Once ${\cal A}_{2,\text{prop}}$  is projected in the $S$-waves for the initial and final pion pairs, Eq.~\eqref{eq:angularprojectionpipis}, we end with the contribution $W_{2,\text{prop}}$ to the kernel $W(s,s,q^2)$ in Eq.~\eqref{eq:resummation}. Keeping in this resummation only terms  up to one-loop, and hence proportional to $G(s)$, one obtains $T_{S2,\text{prop}}$ given by
\begin{align}\label{uni.ex}
T_{S2,\text{prop}} &= -2 \xi_2(s) G(s)  W_{2,\text{prop}}(s,s,q^2) ~\text{,}
\end{align}
where the expansion 
\begin{align}\label{d.exp}
\frac{1}{(1+\xi_2(s) G(s))^2}  =& \sum_{n=0}^\infty (-1)^n (n+1)\, (\xi_2(s)G(s))^n ~\text{,} 
\end{align}
is employed.\footnote{Recall that $V(s)$ in Eq.~\eqref{eq:resummation} is  $\xi_2(s)$ because we are unitarizing a one-loop ChPT calculation for $\pi\pi\, H\to \pi\pi$.} 

However, the result of the one loop calculation in ChPT of the diagram VIII, once properly projected in isospin and $S$-waves as discussed above, gives half of the amplitude in Eq.~\eqref{uni.ex}. Whence Eq.~\eqref{eq:resummation}  is double counting this kind of terms at the one-loop level. Analogously, it can be seen in the $n$-loop terms of the resummation that the contribution of the kernel proportional to $W_{2,\text{prop}}$ is counted $n+1$ times, Eq.~\eqref{d.exp}. This is so because we are missing the proper combinatoric factors as an on-shell factorization scheme for unitarizing is employed. Thus, instead of resumming these terms with $1/(1+VG)^2$, they should be resummed with just $1/(1+VG)$ in order to give the proper diagram counting. Notice that in this case they do not contribute to the double-pole term needed for $F_\sigma(q^2)$, as can be seen from Eqs.~\eqref{eq:def_ff_dp} and Eq.~\eqref{eq:def_ff}. This is also shown schematically in Fig.~\ref{fig:notformfactor}. 
  Had we considered an integral equation for the resummation procedure instead, this kind of contributions would be integrated  giving terms proportional to the three-point function $C_0(s,s',q^2)$, in which the scalar source interacts with intermediate pions, like the terms appearing in the diagram X of Fig.~\ref{fig:pipis_def}. This is not a shortcoming of our approach, because this kind of diagrams are properly included when the kernel $W(s,s',q^2)$ is calculated at higher orders in the chiral counting, as can be seen in Fig.~\ref{fig:pipis_def}. 
\textit{E.g.} at the one-loop level calculation of $W(s,s',q^2)$ one has the diagram X of Fig.~\ref{fig:pipis_def}, that arises from iterating once the pion-propagator contributions at tree level. 

From the previous discussion we remove the terms of the amplitudes with the external scalar source coupled to initial or final asymptotic pion legs from the kernel $W(s,s',q^2)$ in  Eq.~\eqref{eq:resummation}, as  they do not contribute to the scalar form factor of the $\sigma$. The latter requires the coupling of the external scalar source to intermediate pions and vertices.  Now the question arises of how to remove properly the terms arising from the Feynman diagrams with the scalar source attached to  a pion propagator in an external pion leg.  We cannot simply drop these diagrams  because the pion propagator between the source coupling and a pure pionic vertex in an external pion leg may be cancelled by  off-shell terms from the $\pi\pi$ interaction vertex \cite{npa}. Indeed, such contributions  are required in order to have results independent of pion field redefinitions that mix diagrams with different number of pion propagators. 
 Rather, a procedure based on the full on-shell amplitude calculated in ChPT up to some order, which is independent of the former redefinitions, must be given.

 Let us consider the general case, and write these contributions as:
\begin{equation}\label{eq:tconpolo}
 \frac{f(x,y)}{x-x_0}~\text{,}
\end{equation}
where $x=\cos\tilde{\theta}$ and $x_0 = q^2/(4|\vp||\vq|)$.\footnote{We are considering again the case in which the scalar source is attached to $\pi(p_1)$, since the argument for the the other cases is analogous.} Here we have collected in $y$ the rest of the variables. In order to subtract the pure pole contribution in Eq.~\eqref{eq:tconpolo} we subtract from the numerator above the residue of the pole,
\begin{align}
\label{t.sub}
 \frac{f(x,y) - f(x_0,y)}{x-x_0}~.
\end{align}
In the LO case, in view of Eq.~\eqref{eq:TSLO}, this amounts to removing the whole term proportional to the propagator, since it just depends  on $s$ (or $s'$) and not on $\tilde{\theta}$ (or $\theta$), that is, $\partial f(x,y)/\partial x=0$. This subtraction procedure is independent of pion field redefinition because in $f(x_0,y)$ all the pion lines are put on-shell so it cannot contain any  off-shell remainder that could be counterbalanced by other off-shell parts coming from other vertices, and giving rise to possible pion field redefinition dependences.

With this procedure we are then ready to calculate $F_\sigma(q^2)$. For that we define the new amplitude ${\cal B}(s,s',q^2,\theta,\tilde{\theta},\phi,\tilde{\phi})$ obtained from the original ${\cal A}(s,s',q^2,\theta,\tilde{\theta},\phi,\tilde{\phi})$, Eq.~\eqref{amp.def}, by removing the contributions with the scalar source attached to an external pion leg, following the procedure in Eq.~\eqref{t.sub}. In terms of the former we calculate its angular projection as in Eq.~\eqref{eq:angularprojectionpipis}, obtaining the new amplitude $\Phi(s,s',q^2)$:
\begin{align}
\label{phidef}
\Phi(s,s',q^2)&=\frac{1}{32\pi^2}\int \text{d}^2 \Omega\ \mathcal{B}(s,s',q^2,\theta,\tilde{\theta},\phi,\tilde{\phi})~\text{.}
\end{align}
Then, the final expression for the interaction kernel, that we now denote by ${\cal W}(s,s',q^2)$, is ({\it cf.} Eq.~\eqref{exp.wn})
\begin{align}
\label{ws}
{\cal W}&={\cal W}_2+{\cal W}_4~,\nn\\
{\cal W}_2&=\Phi_2~,\nn\\
{\cal W}_4&=\Phi_4+ \Phi_2\,\xi_2(s)G(s) + \Phi_2\,\xi_2(s')G(s')~,
\end{align}
with the subscripts indicating the chiral order as usual.

 The scalar form factor of the $\sigma$ is finally given by
\begin{align}\label{eq:sff_final}
F_\sigma(q^2)=\frac{g_\sigma^2}{V(s_\sigma)^2}{\cal W}(s_\sigma,s_\sigma,q^2)~.
\end{align}

For definiteness let us explicitly give the expressions at LO and NLO for $F_\sigma(q^2)$ from the previous equation:
\begin{align}
F^{{\rm LO}}_\sigma(q^2)&=\frac{(g_\sigma^{{\rm LO}})^2}{V_2(s_\sigma)^2}{\cal W}_2(s_\sigma,s_\sigma,q^2)~,\nn\\
F^{{\rm NLO}}_\sigma(q^2)&=\frac{(g_\sigma^{{\rm NLO}})^2}{\left(V_2(s_\sigma)+V_4(s_\sigma)\right)^2} \nn\\
& \times \left({\cal W}_2(s_\sigma,s_\sigma,q^2)+{\cal W}_4(s_\sigma,s_\sigma,q^2)\right)~.
\end{align}
With 
\begin{align}
\left(g_\sigma^{{\rm LO}}\right)^2&=\lim_{s\to s_\sigma^{{\rm LO}}}(s_\sigma^{{\rm LO}}-s)\frac{V_2(s)}{1+V_2(s) G(s)}~,\nn\\
\left(g_\sigma^{{\rm NLO}}\right)^2&=\lim_{s\to s_\sigma^{{\rm NLO}}}(s_\sigma^{{\rm NLO}}-s)\frac{V_2(s)+V_4(s)}{1+\left(V_2(s)+V_4(s)\right) G(s)}~,
\end{align}
where $s_\sigma^{{\rm LO}}$ and $s_\sigma^{{\rm NLO}}$ are the $\sigma$ pole positions at LO and NLO, respectively, given in Table~\ref{tab:fits_pred}, and for $V_2$ and $V_4$ see  Eq.~\eqref{eq:v4}.

One technical detail is in order. The $\sigma$ resonance is a pole in the second Riemann sheet of $\pi\pi$ scattering for the physical pion mass. As we have seen in Sec.~\ref{sigma.mpi} when increasing the pion mass above some value  the $\sigma$ meson  becomes a bound state and moves into the first Riemann sheet (the corresponding pion mass value is indicated by the arrows in Fig.~\ref{fig:comp_a_b}). Then, Eq.~\eqref{eq:sff_final} has to be understood in the same Riemann sheet as the $\sigma$ pole happens. This requires the evaluation of ${\cal W}(s,s',q^2)$ in Eq.~\eqref{ws} either in the first or second Riemann, according to the value taken for the pion mass.\footnote{This qualification is only relevant for ${\cal W}_4(s,s,q^2)$.} 

We now discuss the analytical continuation of the loop function $C_0(s,s,q^2)$ to the second Riemann sheet (we take from the beginning in the present discussion that $s'=s$), where it is denoted by $C_{0;II}(s,s,q^2)$. The function $C_0(s,s',q^2)$  corresponds to the  three-point one-loop function of diagram X in Fig.~\ref{fig:pipis_def} and its calculation is discussed in Appendix \ref{app:loop}. In order to proceed with the analytical continuation we first evaluate the difference
\begin{align}
\Delta C(s,q^2)=C_0(s+i\epsilon,s+i\epsilon,q^2)-C_0(s-i\epsilon,s-i\epsilon,q^2)
\label{ccut}
\end{align}
for $s$ and $q^2$ real and $q^2<4 M_\pi^2$.\footnote{For $q^2>4 M_\pi^2$ the opening of the $2\pi$ production process introduces additional complications that we skip now since we are mostly interested to values of $q^2$ around zero,  used below to calculate the quadratic scalar radius of the $\sigma$ resonance. The whole region $q^2<4 M_\pi^2$ is of interest and considered by us as well.}
  The second Riemann sheet in $\pi\pi$ scattering is reached by crossing the real $s$-axis above threshold, $s>4 M_\pi^2$, and so we have to consider Eq.~\eqref{ccut} for the same values of $s$. It turns out that a cut in $s$ extends for  $s>2 M_\pi^2 + M_\pi\sqrt{4 M_\pi^2-q^2}\equiv s_{{\rm rc}}$  for which   $\Delta C(s,q^2)$ is non-zero (the same expression for the cut also occurs for 
$s<2 M_\pi^2 - M_\pi\sqrt{4 M_\pi^2-q^2}$). When $q^2\to 0^+$ (this limits gives the same value for the quadratic scalar radius as $q^2\to 0^-$) to cross the real axis for $s>4 M_\pi^2$ implies to consider $ \Delta C(s,q^2)$ given by 
the mentioned cut for $C_0$, $s>s_{{\rm rc}}$, corresponding to $\Delta_b C_0$ in Eq.~\eqref{deltab}. Once this discontinuity is evaluated we continue it analytically in $s$ and $q^2$ and subtract it to $C_0(s,s,q^2)$ (calculated in the first Riemann sheet), as done above to determine $G_{II}(s)$, Eq.~\eqref{eq:gii}. It results,
\begin{align}
C_{0;II}(s,q^2)=C_0(s,q^2)-\Delta C(s,q^2)~.
\end{align}
Notice that for calculating ${\cal W}_4$, Eq.~\eqref{ws}, it is not necessary to use $G_{II}(s)$ when the $\sigma$ pole remains in the second Riemann sheet. This is due to the fact that $\Phi_4$ contains the two-point one-loop function $B_0(s)$, evaluated in Appendix \ref{app:loop}, so that the discontinuity when crossing the unitarity cut above threshold cancels mutually between these two functions. 

\begin{figure}\centering
\includegraphics[height=5.1cm,keepaspectratio]{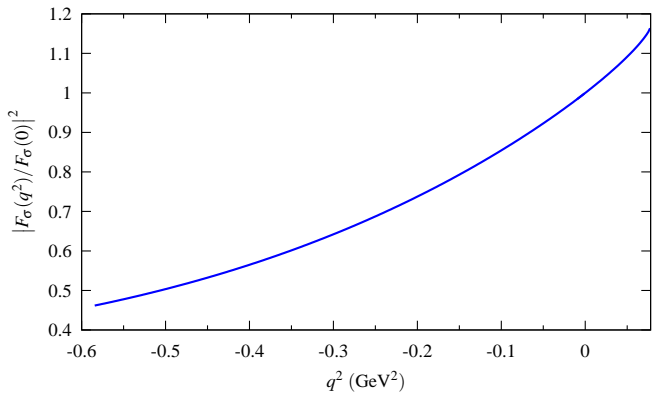}
\caption{The normalized scalar form factor of the $\sigma$ meson calculated at NLO for the physical case with $\sqrt{s_\sigma}$ given  in Table~\ref{tab:fits_pred}. The range in $q^2$ extends from $q^2 \simeq -0.6\ \text{GeV}^2 \simeq -30 M_\pi^2$ up to $q^2 \simeq 0.08\ \text{GeV}^2 \simeq 4M_\pi^2$.\label{fig:plotff}}
\end{figure}

We show in Fig.~\ref{fig:plotff} the modulus squared of $F_\sigma(q^2)$ normalized to $F_\sigma(0)$ for $q^2<4 M_\pi^2$ calculated at NLO with the physical value of $M_\pi$. We observe a monotonous increasing function with $q^2$. The LO result is just a constant because $\Phi_2$ is so and is not shown in the figure (it would be just 1).

\section{Quadratic scalar radius of the $\sigma$ meson and the Feynman-Hellman theorem}\label{sec:r2FH}

The quadratic scalar radius of the $\sigma$ resonance, $\la r^2\ra^\sigma_s$, is related to the scalar form factor of the $\sigma$ by a Taylor expansion around $q^2=0$, 
\begin{align}
F_\sigma(q^2) & = F_\sigma(0) + \left.\frac{\partial F_\sigma(q^2)}{\partial q^2} \right\rvert_{q^2=0}\!\!\!\!\!\! q^2 + \cdots \nn\\
& = F_\sigma(0) \left( 1 + \frac{q^2}{6}\la r^2\ra^\sigma_s  + \cdots \right)~\text{,}
\end{align}
where the ellipsis indicate higher powers of $q^2$ in the Taylor expansion. In this way, 
\begin{equation}
\la r^2\ra^\sigma_s = \frac{6}{F_\sigma(0)}\left.\frac{\partial F_\sigma(q^2)}{\partial q^2} \right\rvert_{q^2=0}~\text{.}
\end{equation}

\begin{figure}
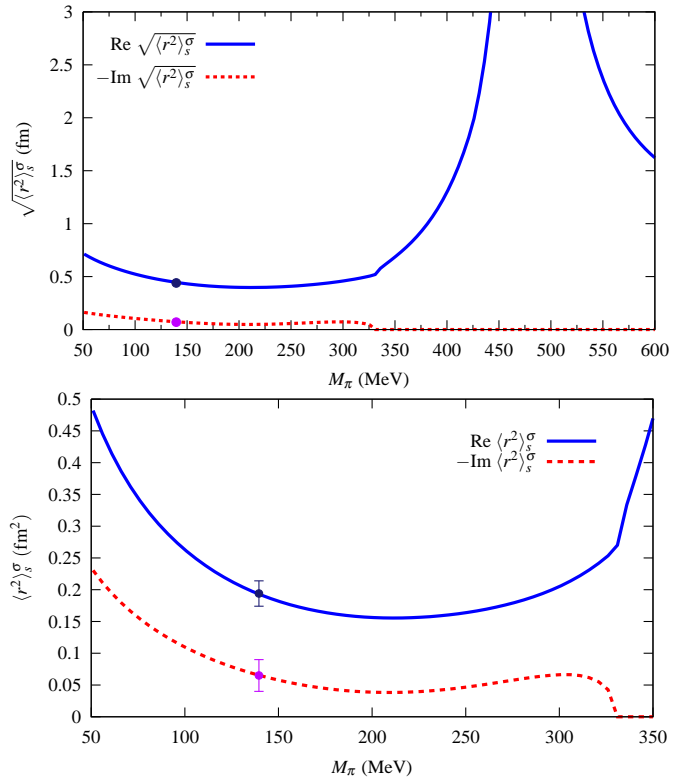

\includegraphics[height=5.1cm,keepaspectratio]{figs_radio_5.mps}
\includegraphics[height=5.1cm,keepaspectratio]{figs_radio_4.mps}
\caption{Top: the square root of the quadratic scalar radius of the $\sigma$ as a function of $M_\pi$ is shown for $0<M_\pi<600\ \text{MeV}$. Bottom: the quadratic scalar radius is represented in the range $0 < M_\pi < 350\ \text{MeV}$. In both panels, the  (blue) solid lines represent the real part of each quantity, whereas the (red) dashed  line is the imaginary part. The points over the curves represent our results for the physical case with their statistical errors, Eq.~\eqref{rs.value}. Due to the scale used they cannot be appreciated in the upper panel.
\label{fig:rss}}
\end{figure}

Notice that, since the form factor reduces to a constant (independent of $q^2$) at LO, we find that $\la r^2\ra^\sigma_s=0$ for this case, similarly as the case for the quadratic scalar radius of the pion \cite{Gasser:1983yg} within standard ChPT. Whence, the quadratic scalar radius must be calculated at least at NLO. Before discussing the results for the physical pion mass case, we study the dependence of $\la r^2\ra^\sigma_s$ with the pion mass. We show the square root of the quadratic scalar radius of the $\sigma$, $\sqrt{\la r^2\ra^\sigma_s}$, in the upper panel of Fig.~\ref{fig:rss} as a function of $M_\pi$, with its real part given by the (blue) solid line and its imaginary part by the (red) dashed line. It diverges in the chiral limit ($M_\pi=0$) and  where the $\sigma$ pole coincides with the two-pion threshold (indicated by the rightmost arrow in Fig.~\ref{fig:evol_m_sig}). The latter point corresponds to a zero energy bound state and as such it must have infinite size, as dictated by elementary quantum mechanics. On the other hand, in the chiral limit $\la r^2\ra^\sigma_s$ also diverges as $\log M_\pi$, similarly as the quadratic scalar or vector  radius of the pion \cite{Gasser:1983yg}, because  the infinite size of the pion cloud around the bosons.  It is relevant to note that the imaginary part of this quantity, despite the $\sigma$ meson has a width larger than 200~MeV for pion masses up to around 300~MeV, as shown in Fig.~\ref{fig:ch_lim}, is much smaller than its real part, which makes its interpretation easier. In the lower panel of the same figure we depict the real (blue solid line) and imaginary (red dashed line) parts of the quadratic scalar radius of the $ \sigma$, $\la r^2 \ra^\sigma_s$. It is notorious that in most of this range of pion mass values the square root of $\la r^2\ra_s^\sigma$ is around 0.5~fm quite independently of the width of the $\sigma$ meson. 

For the physical pion mass we find the values
\begin{align}\label{rs.value}
\la r^2 \ra^\sigma_s= (0.19\pm 0.02) - i\,(0.06 \pm 0.02)\ \mathrm{fm}^2,\nn\\
\sqrt{\la r^2 \ra^\sigma_s}=(0.44 \pm 0.03) - i\,(0.07\pm 0.03)\ \mathrm{fm}~,
\end{align}
with the errors calculated as explained in Sec.~\ref{sec:pipiscattering}. This value is almost the same as the corresponding quadratic scalar radius for $K\pi$, $\la r^2\ra_s^{K\pi}=0.1806\pm0.0049$~fm$^2$ \cite{jamin06}, for which the scalar resonance $\kappa$ (or $K^*_0(800)$), tightly related to the $\sigma$ resonance by $SU(3)$ symmetry \cite{nd,mixing,jaffe1,beveren,black}, plays a leading role \cite{jaminff,jamin}. For comparison, the quadratic scalar radius of the pion is $\la r^2\ra^\pi_s=0.65\pm 0.05$~fm$^2$ \cite{sroca}.\footnote{A recent lattice QCD determination \cite{JLQCDTWQCD} gives $\la r^2 \ra_{s}^\pi = 0.617 \pm 0.079 \pm 0.066\ \text{fm}^2$, or, adding the errors in quadrature, $\la r^2 \ra_{s}^\pi = 0.6 \pm 0.1\ \text{fm}^2$, in good agreement with the value given in Ref.~\cite{sroca}.} It is notorious that the value determined for the scalar radius of the $\sigma$ resonance is  smaller than that for the pion. It is even smaller than the measured quadratic electromagnetic pion radius, $\la r^2\ra_V^{\pi^{\pm}}=0.439\pm 0.008$~fm$^2$ \cite{amendolia86}. However, $\la r^2\ra_s^\sigma$ is similar to the measured  $K^{\pm}$ quadratic charge radius \cite{dally92}, $\la r^2\ra_V^{K^{\pm}}=0.28\pm 0.07$~fm$^2$. Scalar glueballs are expected to have even smaller sizes, $0.1$--$0.2$~fm \cite{hou01,liu92}.

 The value obtained for $ \la r^2\ra_s^\sigma$ in Eq.~\eqref{rs.value} implies that the two scalar isoscalar pions generating the $\sigma$ resonance are so tightly packed that the $\sigma$ resonance becomes a compact state. Whether the two pairs of color singlet  valence quarks $\bar{q}q$  in the two-pion state recombine giving rise to combinations of other possible QCD states as \textit{e.g.}  $q^2\bar{q}^2$ \cite{jaffe1,jaffe2,bob}, glueball, etc is beyond the scope of our  study based on hadronic degrees of freedom. In this respect  the large $N_C$ evolution of the $\sigma$ pole position \cite{nd,pelanc,rios,zhou2010,guo11prd,guo12} is enlightening and clearly indicates that the $\sigma$ resonance is
 not dominantly a glueball or a $\bar{q}q$ resonance. In Refs.~\cite{guo11prd,guo12} it was found that this large $N_C$ behavior is compatible with the fact that this resonance owes its origin to $\pi\pi$ interactions becoming a $\pi\pi$ resonance. This large $N_C$ behavior is also compatible with 
 a $(q\bar{q})^2$ state that fades away as two $q\bar{q}$ mesons as expected in the large $N_C$ limit \cite{manohar}.  
 This picture on the dynamical generation from  $\pi\pi$ interactions of the $\sigma$ meson is also supported by the nontrivial simultaneous fulfillment \cite{guo12} of semi-local duality \cite{pelapeni,guo12} and scalar, pseudoscalar spectral function sum rules \cite{guo12}, both for $N_C=3$ and varying $N_C$.

 On the other hand, for larger values of $M_\pi$, the $\sigma$ meson closely follows the $2\pi$ threshold, as demonstrated in the previous section, and its size is then large. Thus,  in this range of pion masses, the $\sigma$ meson progressively becomes a two-pion molecule and its nature is then much more clear and simple (for $M_\pi\gtrsim 400$~MeV it follows from Fig.~\ref{fig:rss} that $\sqrt{\la r^2\ra}>1.5$~fm).\footnote{A similar value was obtained for the size of the $\Lambda(1405)$ resonance in Ref.~\cite{jidoradius}, which is also a resonance that qualifies as dynamically generated form the meson-baryon interactions \cite{dalitz,kaiserweise,osetramos,plb,jidoset}. In Ref.~\cite{osetnieves} the matter or scalar form factor for this resonance was studied.} This can also be related  from the behavior of the quantity $g^2 dG/ds$ evaluated at $s=s_\sigma$ (and $G$ evaluated in the Riemann sheet in which the pole appears). This quantity is close to one for a composite meson \cite{Weinberg:1965zz,Baru:2003qq,Gamermann:2009uq,Hanhart:2010wh,Hyodo:2011qc,Aceti:2012dd}. We have checked that for the large values of $M_\pi$ in which the $\sigma$ meson is a bound state, we have $g^2 dG/ds \gtrsim 0.8$, which points to a molecular nature. For values of the pion mass close to the physical one, we have instead $g^2 dG/ds \simeq 0$.

\begin{figure}
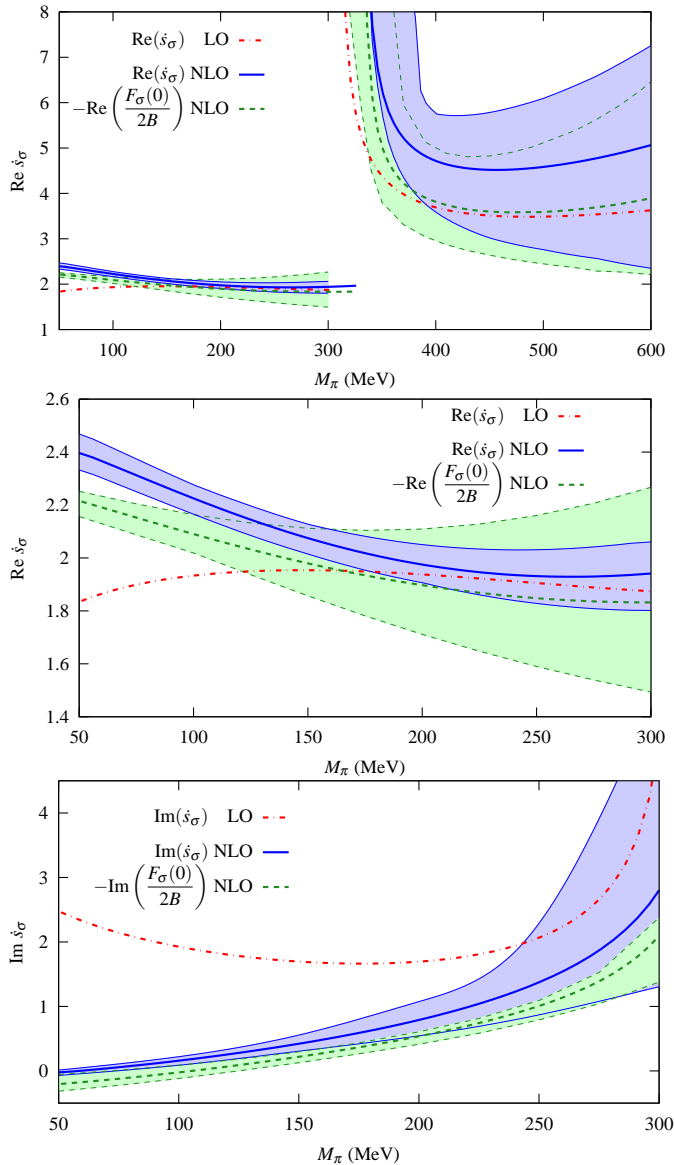

\includegraphics[height=5.1cm,keepaspectratio]{figs_radio_1.mps}
\includegraphics[height=5.1cm,keepaspectratio]{figs_radio_0.mps}
\includegraphics[height=5.1cm,keepaspectratio]{figs_radio_2.mps}
\caption{Feynman-Hellmann theorem: comparison between $ds_\sigma/dM^2$ and $-F_\sigma(0)/2B$, Eq.~\eqref{dsdm2}, as a function of the pion mass. 
 The (blue) thick solid  lines correspond to $ds_\sigma/dM^2$ at NLO, Eq.~\eqref{eq:dssig}, whereas the (red) dot-dashed  lines are evaluated at LO, where $ds_\sigma/dM^2 = ds_\sigma/dM_\pi^2$. The (green) dashed  lines are $-F_\sigma(0)/2B$. From top to bottom, in the first panel the real part of the quantities are represented in the range  $M_\pi$, $50 < M_\pi < 600\ \text{MeV}$. In the second panel, the same is shown for $50 < M_\pi < 300\ \text{MeV}$. The bottom panel shows their imaginary part. 
\label{fig:fs0}}
\end{figure}

Another interesting point is to consider the relation between $F_\sigma(0)$ and the derivative of the $\sigma$ pole with respect to the quark mass. According to the Feynman-Hellmann theorem \cite{fh}, one has the relation:
\begin{align}
\label{dsdm2.fs}
\frac{ds_\sigma}{dM^2}&=-\frac{F_\sigma(0)}{2B}~.
\end{align}
Notice that $F_\sigma(0)$ is proportional to $B$ and precisely their ratio is not ambiguous. On the other hand, $ds_\sigma/dM_\pi^2$ is given 
in Eq.~\eqref{eq:dssig}. Then we can write:
\begin{align}
\frac{d s_\sigma}{d M^2}=& -\frac{g_\sigma^2(M_\pi^2)}{V(s_\sigma,M_\pi^2)^2}\left( \dot{V}(s_\sigma,M_\pi^2) - V(s_\sigma,M_\pi^2)^2 
\dot{G}_{II}(s_\sigma,M_\pi^2)  \right)\nn\\
\times & \frac{dM_\pi^2}{dM^2}~.
\label{dsdm2}
\end{align}
The dependence of $M_\pi^2$ on $M^2$ is worked out up to ${\cal O}(M_\pi^4)$ in Eq.~\eqref{mpi2} from where one obtains:
\begin{align}
\frac{d M_\pi^2}{d M^2}&=1-\frac{M_\pi^2}{16 \pi^2 F_\pi^2}\left(\bar{l}_3 - \frac{1}{2} \right) +{\cal O}(M_\pi^4)~.
\end{align}
We show our results for $-F_\sigma(0)/2B$ at NLO and compare them with $ds_\sigma/dM^2$ in Fig.~\ref{fig:fs0}, so as to check Eq.~\eqref{dsdm2.fs}. 
 In the upper two panels we show the real part and in the bottom one the imaginary part. The agreement is certainly remarkable for $M_\pi\lesssim 300$~MeV, at the level of just a few percents of difference. This range of pion masses is highlighted in the second and third panels, from top to bottom. Let us note that in Eq.~\eqref{dsdm2.fs} we are comparing two quantities that are obtained from the chiral expansion of two different interacting kernels. The expansion is not performed on the full amplitudes and this is why there is not a perfect agreement, as it is the case in the standard perturbative calculations of ChPT \cite{Gasser:1983yg,chptsu3}. In our case the factor $V(s_\sigma,M_\pi^2)^2$ multiplying $\dot{G}_{II}(s_\sigma,M_\pi^2)$ in the right-hand side of Eq.~\eqref{dsdm2} is equal to $(V_2+V_4)^2$, while $\Phi_4$ from Eq.~\eqref{phidef} only contains $V_2^2$, because it is a ChPT one-loop calculation at ${\cal O}(p^4)$.\footnote{Notice that the derivative with respect to $M_\pi^2$ of the function $G(s,M_\pi^2)$ is proportional to $C_0(s,s,M_\pi^2)$.} Thus, the differences correspond to higher order terms in the calculation of $F_\sigma(0)$, beyond the ${\cal O}(p^4)$ or NLO calculation of the kernel ${\cal W}(s,s',q^2)$, Eq.~\eqref{ws}, performed in the present work. 

Another point also worth mentioning is the fact that the left-hand side of Eq.~\eqref{dsdm2} does not involve any contribution with pion propagators in the external legs but the derivative acts on the vertices and intermediate pion propagators in loop functions. This is also the case in $F_\sigma(0)$ once the pion propagators in the external legs are removed as explained in Sec.~\ref{subsec:formfactor}.

\section{Summary and conclusions}\label{sec:conclusions}

  In this work we have discussed the nature of the $\sigma$ resonance (nowadays also called $f_0(600)$ in the PDG \cite{pdg}) by evaluating its quadratic scalar radius, $\la r^2\ra_s^\sigma$. This allows one to have a quantitative idea of the size of this resonance.

There are many studies since the nineties based on supplementing Chiral Perturbation Theory with non-perturbative $S$-matrix methods, that  clearly indicate  a dynamical origin for the $\sigma$  resonance due to the isoscalar scalar $\pi\pi$ strong self interactions \cite{tr1,e3pi.tr,prl.tr,dobado,pela1,npa,prl}. More recent studies based on the dependence with $N_C$ of the $\sigma$ pole \cite{nd,pelanc,rios,guo11prd,arriola1} also corroborate that this resonance cannot be qualified as a purely $\bar{q}q$ or glueball, with the pole trajectories compatible with the expectations for a meson-meson dynamically generated object or a four-quark state. In the large $N_C$ limit it is well known that loops are suppressed  so that  the $\pi\pi$ rescattering vanishes away and then the $\sigma$ resonance pole disappears according to Refs.~\cite{nd,guo11prd,guo12}. These results have been strongly supported recently \cite{guo12} by the simultaneous fulfillment of semi-local duality \cite{pelapeni,guo12} and scalar, pseudoscalar spectral sum rules \cite{guo12}, both for $N_C=3$ and varying $N_C$.

The next question is whether the two pions are loosely distributed, so that the $\sigma$ meson might be qualified as  molecular or, on the contrary, they overlap each other giving rise to a compact object of a  size comparable or even smaller than that of its constituents. A proper way to answer this question is to determine quantitatively the size of the $\sigma$ resonance. For that we calculate in this work the quadratic scalar radius of this resonance 
obtaining the value  $\la r^2\ra_s^\sigma=(0.19\pm 0.02)-i\,(0.06\pm 0.02)$ fm$^2$. Despite the $\sigma$ has a large width the resulting value for the quadratic scalar radius is almost a real quantity, which makes easier its physical interpretation. This value is very close to the $K\pi$ quadratic scalar radius, $\la r^2\ra_s^{K\pi}=0.1806\pm0.0049$~fm$^2$ \cite{jamin06}, similar to the measured $K^{\pm}$ quadratic charge radius \cite{dally92}, $\la r^2\ra_V^{K^{\pm}}=0.28\pm 0.07$~fm$^2$, and smaller than the quadratic scalar radius of the pion, $\la r^2\ra_s^\pi=0.65\pm 0.05$ fm$^2$ \cite{sroca}. This means that the $\sigma$ is certainly a compact object. The square root of its quadratic scalar radius is $\sqrt{\la r^2\ra_s^\sigma}=(0.44 \pm 0.03) - i\, (0.07\pm 0.03)$~fm. 

We have further tested our result for the size of the $\sigma$ by considering the dependence of $\la r^2\ra_s^\sigma$ on the pion mass. As $M_\pi$ rises the $\sigma$ meson mass follows the $2\pi$ threshold. This fact has been recently observed in the lattice QCD calculation of Ref.~\cite{Prelovsek:2010kg}, and was pointed out much earlier in Refs.~\cite{nd,mixing} as well as in the more recent work \cite{Pelaez:2010fj}. In such situation, with a small binding energy, the expected size of the $\sigma$ resonance should be definitely larger than that of a hadron.  We obtain  a quadratic scalar radius  that increases rapidly as soon as the width of the $\sigma$ meson tends to vanish, which for our present NLO fit occurs for pion masses above $\simeq 330$~MeV. In this way, already for pion masses around 370~MeV, $\sqrt{\la r^2\ra_s^\sigma}$ is larger than 1~fm and diverges for $M_\pi\simeq 470$~MeV, precisely the value at which  the $\sigma$ resonance becomes a zero binding energy bound state. In this case, a molecular or $\pi\pi$ bound state image is appropriate for the $\sigma$ meson. For even higher pion masses, the binding energy is still  small which gives rise to large sizes for the $\sigma$. Nevertheless, we observe a steady (albeit weak) tendency to increase the binding energy for higher pion masses so that its size tends to dismiss progressively, but for the mass range explored in this work it is always $\gtrsim 1.5$~fm. The clear tendency of the $ \sigma$ resonance to follow the two-pion threshold is a manifest  indication for this resonance being a meson-meson dynamically generated one. For smaller pion masses between 50 and 300~MeV the square root of the quadratic scalar radius of the $\sigma$ meson is rather stable with a value around 0.5~fm, independently of its width. 

The value of the scalar form factor of the $\sigma$ resonance at $q^2=0$, $F_\sigma(0)$, is related via the Feynman-Hellmann theorem with the derivative of the $\sigma$ pole position with respect to the pion mass. Within uncertainties, we have checked the fulfillment of such relation.

We have studied $\pi\pi$ scattering in NLO $SU(2)$ Unitary Chiral Perturbation Theory as well. We obtain a good reproduction of $\pi\pi$ phase shifts for $I=0$ and $I=2$, and also for lattice QCD results of the $I=2$ scattering length $a_0^2$ and $F_\pi$. We have offered a detailed comparison between different precise determinations in the literature, including our present calculation, of the $\sigma$ meson mass and width, and of the threshold parameters $a_0^0$, $b_0^0$. The resulting average values are $a_0^0=0.220\pm 0.003$ and $b_0^0 M_\pi^2=0.279\pm 0.003$. For the $\sigma$ meson pole parameters we take the mean of the different values with the result $M_\sigma=458\pm 14$~MeV and $\Gamma_\sigma/2=261\pm 17$~MeV. Our own determinations obtained here at NLO in Unitary ChPT are  $a_0^0 = 0.219 \pm 0.005$, $b_0^0 M_\pi^2 = 0.281 \pm 0.006$, $M_\sigma = 440 \pm 10\ \text{MeV}$ and $\Gamma_\sigma/2=238 \pm 10$~MeV. 

\acknowledgements
JAO would like to thank Eulogio Oset for an interesting discussion. We also are pleased to Guillermo R\'{\i}os for having provided us with the IAM results of Fig.~\ref{fig:ch_lim}. This work is partially funded by the grants MICINN FPA2010-17806 and the Fundaci\'on S\'eneca 11871/PI/09. We also thank the financial support from the BMBF grant 06BN411, the EU-Research Infrastructure Integrating Activity ``Study of Strongly Interacting Matter'' (HadronPhysics2, grant n. 227431) under the Seventh Framework Program of EU and the Consolider-Ingenio 2010 Programme CPAN (CSD2007-00042).

\appendix
\section{Loop functions}\label{app:loop}
\begin{figure}
\includegraphics[width=0.45\textwidth,keepaspectratio]{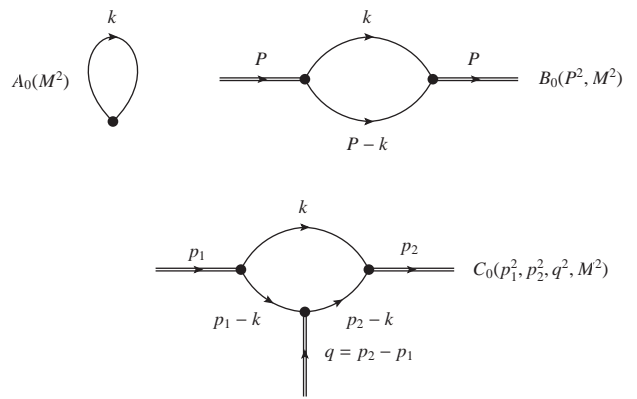}
\caption{Diagrams for the one-loop functions $A_0$, $B_0$ and $C_0$ (from left to right and top to bottom, respectively).\label{fig:loopfun}}
\end{figure}

\newcommand{\comint}{i\!\!\! \int \!\!\! \frac{d^4 k}{(2\pi)^4}}
\newcommand{\ie}{i \epsilon}
\newcommand{\rlogc}{R + \log\frac{M^2}{\mu^2}}
In this Appendix, we give the loop functions used through the paper. We start with the scalar one-, two- and three-point one-loop integrals, denoted by $A_0$, $B_0$ and $C_0$, respectively, and depicted in Fig.~\ref{fig:loopfun}. The vector and tensor integrals, defined later, can be cast in terms of the former. Special attention is dedicated to the case of the three-point function,  whose cuts are also calculated since they are needed in order to evaluate the scalar form factor of the $\sigma$ meson. Notice that  all the internal masses are equal, as we only have pions as degrees of freedom. For this reason, we do not include the dependence on the internal mass $M^2$ in the following (except for the case of the function $A_0$, which does not depend on any external momenta).

\subsection*{Scalar loop integrals}
The simplest one is the one-point loop integral, given by:
\begin{equation}
A_0(M^2) = \comint \frac{1}{k^2 - M^2 + \ie}~.
\end{equation}
In dimensional regularization, it results:
\begin{equation}\label{eq:loopA0}
A_0(M^2) = \frac{M^2}{16\pi^2} \left( \rlogc \right)~,
\end{equation}
with 
\begin{equation}
R = \mu^{n-4}\left( \frac{2}{n-4} - (1+\Gamma'(1) + \log 4\pi) \right)~,
\end{equation}
and $n \to 4$.

The two-point function is:
\begin{equation}
B_0(P^2) = \comint \frac{1}{\Big(k^2 - M^2 + \ie\Big)\Big((k-P)^2-M^2 + \ie\Big)}~,
\end{equation}
and analogously it is evaluated in dimensional regularization, with the result
\begin{equation}\label{eq:loopB0}
B_0(P^2) = \frac{1}{16\pi^2} \left( \rlogc -1 - \sigma(P^2) \log\frac{\sigma(P^2)-1}{\sigma(P^2)+1} \right)~,
\end{equation}
with $\sigma(P^2) = \sqrt{1-4M^2/P^2}$. Since the function is divergent and $\mu$-dependent, we define the subtracted function, $\bar{B}_0(P^2)$,
\begin{equation}\label{eq:loopB0_sub}
\bar{B}_0(P^2) = B_0(P^2) - \frac{\displaystyle\rlogc}{16\pi^2}~.
\end{equation}
This is the function that will appear in the amplitudes, because the term  subtracted cancels out with the alike terms in the loops and the chiral counterterms. The same procedure, applied to $A_0(M^2)$, gives
\begin{equation}
\bar{A}_0 = A_0(M^2) - \frac{M^2}{16\pi^2} \left( \rlogc \right) = 0~,
\end{equation}
this is why in the amplitudes of Appendix~\ref{app:amp} there is no dependence on $\bar{A}_0$.

The three-point function is defined by:
\begin{align}
C_0(p_1^2,p_2^2,q^2) & = \comint \frac{1}{k^2 - M^2 + \ie} \\
& \times \frac{1}{\Big((k-p_1)^2-M^2 + \ie\Big)\Big((k-p_2)^2-M^2+\ie\Big)}~,\nn
\end{align}
and it depends on the three scalars $p_1^2$, $p_2^2$ and $q^2 \equiv (p_1-p_2)^2$. It is finite and after some manipulations  \cite{'tHooft:1978xw}, it can be cast in the integral form:
\begin{align}\label{eq:c0_int_form}
C_0(p_1^2,p_2^2,q^2) & = \frac{1}{16\pi^2 \lambda(p_1^2,p_2^2,q^2)^{1/2}} \left\{ \vphantom{\int \frac{\log}{t}} \right. \nn  \\
    & \int_0^1 dz \frac{\log f(p_1^2,z)-\log f(p_1^2,z_1)}{z-z_1} +\nn \\
  + & \int_0^1 dz \frac{\log f(p_2^2,z)-\log f(p_2^2,z_2)}{z-z_2} +\nn \\
  + & \int_0^1 dz \frac{\log f(q^2  ,z)-\log f(q^2  ,z_3)}{z-z_3} 
\left. \vphantom{\int \frac{\log}{t}} \right\}~,
\end{align}
where we have defined:
\begin{align}
f(p^2,z) & = p^2 z(z-1) + M^2 - \ie~, \\
\lambda(a,b,c) & = a^2+b^2+c^2-2ab-2bc-2ac~, \\
z_1 & = \frac{1}{2} \left( 1 + \frac{p_1^2-p_2^2-q^2  }{\lambda(p_1^2,p_2^2,q^2)^{1/2}} \right)~,\\
z_2 & = \frac{1}{2} \left( 1 + \frac{p_2^2-p_1^2-q^2  }{\lambda(p_1^2,p_2^2,q^2)^{1/2}} \right)~,\\
z_3 & = \frac{1}{2} \left( 1 + \frac{q^2  -p_1^2-p_2^2}{\lambda(p_1^2,p_2^2,q^2)^{1/2}} \right)~.
\end{align}
The usefulness of Eq.~\eqref{eq:c0_int_form} lies in the fact that it is well suited for its analytical continuation to the complex plane, which is needed in our case, since the cases $p_1^2 = p_2^2 = s_\sigma$ are studied. Notice that the residues of the integrals when $z \to z_i$ are zero because of the form of the numerators, and also that, since $z(z-1) \leqslant 0$ for $z\in [0,1]$, the arguments of the logarithms do not cross any cut.

On the other hand, since the pole of the $\sigma$ resonance appears in the unphysical Riemann sheet, we need to calculate the amplitudes $\pi\pi H \to \pi\pi$ in this sheet. This involves the function $C_0(s,s,q^2)$ in this sheet,\footnote{Notice that we have already taken that $s=s'$, since this will be the case in the $\sigma$ form factor, $s=s'=s_\sigma$.} and this is not so trivial as in the case of the function $G(s)$ (see Eqs.~\eqref{eq:gii} and \eqref{eq:gii_2}). For that purpose, we calculate the discontinuity along the unitarity cut of the function $C_0(s,s,q^2)$:
\begin{equation}
\Delta C_0 = C_0(s+\ie,s+\ie,q^2) - C_0(s-\ie,s-\ie,q^2)~,
\end{equation}
for  $s \geqslant 4M^2$. This can be obtained directly from the integral representation in Eq.~\eqref{eq:c0_int_form}, and the result depends on the value of $q^2$. We are interested mainly in the case $q^2 \leqslant 0$, and we find two cases:
\begin{align}
\label{delta}
\Delta_a C_0 & = \frac{i}{4\pi \lambda(s,s,q^2)^{1/2}} \log \left( \frac{z_- - z_1}{z_+ - z_1}\right)\quad \text{for } q^2 \leqslant q_{an}^2~, \\
\Delta_b C_0 & = \frac{i}{4\pi \lambda(s,s,q^2)^{1/2}} \log \left(\frac{1-z_1}{-z_1} \frac{z_- - z_1}{z_+ - z_1}\right)  \nn \\
& + \frac{i}{8\pi \lambda(s,s,q^2)^{1/2}} \log \left( \frac{1-z_3}{-z_3} \right)\quad \text{for }  q_{an}^2 \leqslant q^2 \leqslant 4M^2~,
\label{deltab}
\end{align}
where we have defined $z_{\pm} = \frac{1}{2} (1 \pm \sigma(s))$. In the previous equation $q_{an}^2$ is the so called anomalous threshold, given by $M^2 q_{an}^2 = -s(s-4M^2) \leqslant 0$, where the last inequality follows from $s \geqslant 4M^2$. The case that connects continuously with  $q^2 = 0$ corresponds to $\Delta_b C_0$, which is the required one  in  the calculation of the quadratic scalar radius. For more details on the analytical extrapolation chosen see the discussion in Sec.~\ref{subsec:formfactor}.  

As a cross-check of the validity of our procedure, let us note that the function $C_0$ is related for $q^2=0$ to the  derivative of the function $G(s)$ with respect to $M^2$, denoted by $d G(s,M^2)/dM^2 = \dot{G}$. Indeed, one has
\begin{equation}
2 C_0(s,s,0) = \dot{G}(s) = \frac{1}{8\pi^ 2 s \sigma(s)} \log \frac{\sigma(s)-1}{\sigma(s)+1}~.
\end{equation}

 In the calculation of the scalar form factor $C_0$  appears, while in the derivative of the $\sigma$ pole position, $\dot{s}_\sigma$, one has  $\dot{G}$. Both  are related through the Feynman-Hellmann theorem, Eq.~\eqref{dsdm2.fs}, and thus, the unphysical Riemann sheet for the function $C_0$ must be related to that of the function $G(s)$. When the pole is in the unphysical Riemann sheet, we have:
\begin{equation}
\dot{G}_{II}(s) = \dot{G}_I(s) - \dot{\Delta} G(s) = \dot{G}_I(s) - \frac{i}{4\pi s \sigma(s)}~.
\end{equation}
If we now calculate the value $\Delta_b C_0(s,s,0)$, we find:
\begin{equation}
\Delta_b C_0(s,s,0) = \frac{i}{8\pi s \sigma(s)}~,
\end{equation}
so that
\begin{align}
C_{0;II}(s,s,0) & = C_{0;I}(s,s,0) - \Delta_b C_0(s,s,0) \nn\\ & = C_{0;I}(s,s,0) - \frac{i}{8\pi s \sigma(s)}~,
\end{align}
which implies, as stated, $2 C_{0;II}(s,s,0) = \dot{G}_{II}(s)$.

\subsection*{Vector and tensor loop integrals}
Vector and tensor loop integrals appear throughout the amplitudes in Appendix \ref{app:amp}. We reduce them to the scalar ones by means of the Passarino-Veltman method \cite{Passarino:1978jh}. We start with the two-point vector and tensor integrals, defined by
\begin{equation}
B_{\{\mu;\mu\nu\}} = \comint \frac{\{k_\mu;k_{\mu}k_{\nu}\}}{\Big(k^2 - M^2 + \ie\Big)\Big((k-P)^2-M^2 + \ie\Big)}~.
\end{equation}
On Lorentz-invariance grounds, we can write
\begin{equation}
B^{\mu} = -P^{\mu} B_{11}(P)~,
\end{equation}
where the minus sign is introduced for convenience, and, performing the contraction $P^\mu B_{\mu}$, it can be shown that:
\begin{equation}
B_{11}(P^2) = - \frac{1}{2}B_0(P^2)~.
\end{equation}
Analogously, the tensor integral can be decomposed as:
\begin{equation}
B_{\mu\nu} = P_\mu P_\nu B_{20}(P^2) + g_{\mu\nu} P^2 B_{21}(P^2)~,
\end{equation}
and the following results, by the appropriate contractions, are obtained:
\begin{align}
P^2 B_{20}(P^2) & = \frac{A_0(M^2)}{3}  - \frac{M^2-P^2}{3} B_0(P^2) + \frac{1}{48\pi^2}\left(M^2-\frac{P^2}{6}\right)~,\\
P^2 B_{21}(P^2) & = \frac{A_0(M^2)}{6}  + \frac{4M^2-P^2}{12} B_0(P^2) - \frac{1}{48\pi^2}\left(M^2-\frac{P^2}{6}\right)~.
\end{align}

For the three-point vector and tensor integrals, we define:
\begin{align}
C_{\{\mu;\mu\nu\}} & = \comint \frac{\{k_\mu;k_{\mu}k_{\nu}\}}{k^2 - M^2 + \ie} \\
& \times \frac{1}{\Big((k-p_1)^2-M^2 + \ie\Big)\Big((k-p_2)^2-M^2+\ie\Big)}\nn~,
\end{align}
and
\begin{align}
C^{\mu} & = - p_1^\mu C_{11} - p_2^\mu C_{12}~, \\
C^{\mu\nu}& = p_1^{\mu}p_1^{\nu} C_{21} + p_2^{\mu}p_2^{\nu} C_{22} + \left( p_1^\mu p_2^\nu + p_1^\nu p_2^\mu \right) C_{23} + g^{\mu\nu} C_{24}~,
\end{align}
where for simplifying the writing  we have omitted the arguments in  $C_{ij}(p_1^2,p_2^2,q^2)$. The results for these functions are:
\begin{align}
C_{11} & = \left( p_2^2 B_0(p_2^2) - p_1 p_2 B_0(p_1^2) - (p_2^2-p_1p_2)B_0(q^2) \right. \nn \\ & \left. - p_2^2(p_1^2-p_1p_2)C_0(p_1^2,p_2^2,q^2) \right)/(2\det {\cal H})~, \\
C_{12} & = \left( p_1^2 B_0(p_1^2) - p_1 p_2 B_0(p_2^2) - (p_1^2-p_1p_2)B_0(q^2) \right. \nn \\ & \left. - p_1^2(p_2^2-p_1p_2)C_0(p_1^2,p_2^2,q^2) \right)/(2\det {\cal H})~, \\
C_{24} & = -\frac{1}{64\pi^2} + \frac{M^2}{2} C_0(p_1^2,p_2^2,q^2) \nn \\
& + \frac{1}{4} \left( B_0(q^2) + p_1^2 C_{11} + p_2^2 C_{12} \right)~, \\
C_{21} & = \left( p_2^2 R_a - p_1 p_2 R_c \right) /\det {\cal H}~, \\
C_{22} & = \left( p_1^2 R_d - p_1 p_2 R_b \right) /\det {\cal H}~, \\
C_{23} & = \left( p_1^2 R_c + p_2^2 R_b -p_1 p_2 (R_a + R_d) \right) / (2\det {\cal H})~,
\end{align}
with
\begin{equation}
{\cal H} = \left( \begin{array}{cc} p_1^2 & p_1 p_2 \\ p_1 p_2 & p_2^2\end{array} \right)~,
\end{equation}
and
\begin{align}
R_a & = \frac{1}{4} B_0(q^2) - \frac{1}{2} p_1^2 C_{11} - C_{24}~, \\
R_b & = \frac{1}{4} B_0(q^2) - \frac{1}{2} p_1^2 C_{12} - \frac{1}{4}B_0(p_2^2)~, \\
R_c & = \frac{1}{4} B_0(q^2) - \frac{1}{2} p_2^2 C_{11}~- \frac{1}{4}B_0(p_1^2), \\
R_d & = \frac{1}{4} B_0(q^2) - \frac{1}{2} p_2^2 C_{12} - C_{24}~.
\end{align}

Analogously to the scalar loop integrals, we define the subtracted functions $\bar{B}_{ij}$ and $\bar{C}_{ij}$ by substituting in their expressions given above $A_0 \to \bar{A}_0$ and $B_0 \to \bar{B}_0$. The amplitudes $\pi\pi\,H\to\pi\pi$ in Appendix~\ref{app:amp} are then written in terms of finite and scale independent functions.


\section{$\pi\pi\,H \to \pi\pi$ amplitudes}\label{app:amp}
\newcommand{\prUN}{D_1}
\newcommand{\prDO}{D_2}
\newcommand{\prTR}{D_3}
\newcommand{\prCU}{D_4}
\newcommand{\tempamp}{}
\newcommand{\hpr}{\hphantom{'}}
\newcommand{\rpr}{'}

In this Appendix, the amplitudes $\pi\pi\,H\to\pi\pi$ are given for completeness. We follow the nomenclature given in Fig.~\ref{fig:pipis_def}. The finite contributions to every amplitude are given once the infinite and scale dependent terms are cancelled among them. In this way, the amplitudes are written in terms of the finite and scale independent constants $\bar{l}_i$ as well as the subtracted loop functions defined in Appendix~\ref{app:loop}, $\bar{B}_0$, etc... The diagrams denoted in Fig.~\ref{fig:pipis_def} by VI, XI and XII, both in the case of $\ppnn$ and $\ppnc$, are proportional to the tadpole function, $A_0(M^2)$, so they do not contribute to the finite amplitude, as explained before.

In the subsequent, unless the opposite is stated, the subscript $i=1,\ldots,4$ indicates the pion leg with four-momentum $p_i$ to which the scalar source is attached. The functions $D_i$, corresponding to the inverse of the pion propagators when the scalar source is attached to the $i_{\text{th}}$ external pion leg, are used through this Appendix. These functions were defined 
in Eq.~\eqref{eq:ds.def}.

\subsection{Diagrams I}
\subsubsection{$\ppnn$}
\begin{align}
T^{\text{\tiny (LO)}}   & =  - \frac{6B}{F_\pi^2}~, \\
T^{\text{\tiny (NLO)}}  & =  - \frac{3B}{F_\pi^4}\frac{M_\pi^2}{4\pi^2}\bar{l}_4~.
\end{align}
\subsubsection{$\ppnc$}
\begin{align}
T^{\text{\tiny (LO)}}   & =  - \frac{2B}{F_\pi^2}~, \\
T^{\text{\tiny (NLO)}}  & =  - \frac{B}{F_\pi^4}\frac{M_\pi^2}{4\pi^2}\bar{l}_4~.
\end{align}
The NLO result corresponds to the LO diagram I multiplied by $2\delta Z$, with the latter given in Eq.~\eqref{mpi2}. In addition $M^2$, $F^2$ are expressed in terms of the physical values according to the expansions of Eqs.~\eqref{mpi2} and \eqref{eq:fpiNLO}.
\subsection{Diagrams II}
\subsubsection{$\ppnn$}
\renewcommand{\tempamp}[2]{T_{#1}^{\text{\tiny (LO)}}   & =   \frac{2B}{F_\pi^2} \left( 1 + \frac{M_\pi^2}{#2} \right)}
\begin{align}
\tempamp{1}{\prUN}~,\nn\\
\tempamp{2}{\prDO}~,\nn\\
\tempamp{3}{\prTR}~,\nn\\
\tempamp{4}{\prCU}~.
\end{align}
\renewcommand{\tempamp}[2]{T_{#1}^{\text{\tiny (NLO)}}   & =   
\frac{B}{F_\pi^4} \left\{\frac{\bar{l}_4M_\pi^2}{4\pi^2} + \frac{M_\pi^4}{#2} \frac{4\bar{l}_4 - 3\bar{l}_3}{16\pi^2}\right\}
}
\begin{align}
\tempamp{1}{\prUN}~,\nn\\
\tempamp{2}{\prDO}~,\nn\\
\tempamp{3}{\prTR}~,\nn\\
\tempamp{4}{\prCU}~.
\end{align}

\subsubsection{$\ppnc$}
\renewcommand{\tempamp}[3]{T_{#1}^{\text{\tiny (LO)}}   & =   \frac{2B}{F_\pi^2}\frac{#2  - M_\pi^2}{#3}}
\begin{align}
\tempamp{1}{s\hphantom{'}}{\prUN}~,\nn\\
\tempamp{2}{s\hphantom{'}}{\prDO}~,\nn\\
\tempamp{3}{s'}{\prTR}~,\nn\\
\tempamp{4}{s'}{\prCU}~.
\end{align}
\renewcommand{\tempamp}[3]{iT_{#1}^{\text{\tiny (NLO)}}   & =   \frac{B}{F_\pi^2}\frac{M_\pi^2}{#3}\frac{4\bar{l}_4(#2-M_\pi^2) - \bar{l}_3 M_\pi^2}{16\pi^2}}
\begin{align}
\tempamp{1}{s\hphantom{'}}{\prUN}~,\nn\\
\tempamp{2}{s\hphantom{'}}{\prDO}~,\nn\\
\tempamp{3}{s'}{\prTR}~,\nn\\
\tempamp{4}{s'}{\prCU}~.
\end{align}
The NLO results are obtained by multiplying the  LO ones  by  $3\delta Z$ and with $M^2$, $F^2$ re-expressed in terms of the physical $M_\pi^2$ and $F_\pi^2$, respectively, according to Eqs.~\eqref{mpi2} and \eqref{eq:fpiNLO}. Notice that in addition to the factor $Z^2$ from the wave function renormalization of the external pion legs there is an extra factor $Z$ from  the renormalized pion propagator, Eq.~\eqref{pp.ren}.

\subsection{Diagrams III}
The diagrams III and higher in numeration are purely NLO contributions. To simplify the writing we then  omit the superscript {\small NLO} in the corresponding amplitudes. 
\subsubsection{$\ppnn$}
\begin{equation}
T = -\frac{3B}{F_\pi^4} \frac{\bar{l}_4}{8\pi^2} q^2~.
\end{equation}
\subsubsection{$\ppnc$}
\begin{equation}
T = -\frac{B}{F_\pi^4} \frac{\bar{l}_4}{8\pi^2} q^2~.
\end{equation}

\subsection{Diagrams IV}
\subsubsection{$\ppnn$}
\renewcommand{\tempamp}[2]{T_{#1} & =  \frac{B}{F_\pi^4} \frac{\bar{l}_4 q^2 - \bar{l}_3 M_\pi^2}{8\pi^2} \left( 1 + \frac{M_\pi^2}{#2} \right)}
\begin{align}
\tempamp{1}{\prUN}~,\nn\\
\tempamp{2}{\prDO}~,\nn\\
\tempamp{3}{\prTR}~,\nn\\
\tempamp{4}{\prCU}~.
\end{align}
\subsubsection{$\ppnc$}
\renewcommand{\tempamp}[3]{T_{#1} & =  \frac{B}{F_\pi^4} \frac{\bar{l}_4 q^2 - \bar{l}_3 M_\pi^2}{8\pi^2} \left( \frac{#2  - M_\pi^2}{#3} \right)}
\begin{align}
\tempamp{1}{s\hphantom{'}}{\prUN}~,\nn\\
\tempamp{2}{s\hphantom{'}}{\prDO}~,\nn\\
\tempamp{3}{s'}{\prTR}~,\nn\\
\tempamp{4}{s'}{\prCU}~.
\end{align}

\subsection{Diagrams V}
\subsubsection{$\ppnn$}
\renewcommand{\tempamp}[5]{T_1 & =  -\frac{B}{F_\pi^4}\frac{\bar{l}_1 + 2\bar{l}_2}{24\pi^2} \left( #5 + 2M_\pi^2 - \frac{#2^2  + #3^2 + #4^2 - 4 M_\pi^4}{#5} \right) }
\begin{align}
\tempamp{1}{s\hpr}{t\rpr}{u\rpr}{\prUN}~,\nn\\
\tempamp{2}{s\hpr}{t\hpr}{u\hpr}{\prDO}~,\nn\\
\tempamp{3}{s\rpr}{t\rpr}{u\hpr}{\prTR}~,\nn\\
\tempamp{4}{s\rpr}{t\hpr}{u\rpr}{\prCU}~.
\end{align}
\subsubsection{$\ppnc$}
In these amplitudes, we define:
\begin{align}
P(s,t,u) & = 4 M_\pi^4(\bar{l}_1 + 2\bar{l}_2) + \bar{l}_1 s^2  + \bar{l}_2 (t^2+ u^2) \nn\\ & - 2M_\pi^2(2 \bar{l}_1 s + 2\bar{l}_2(t + u))
\end{align}
\renewcommand{\tempamp}[5]{T_{#1} =  -\frac{B}{24\pi^2 F_\pi^4} \left(\bar{l}_1 #2    + \bar{l}_2 (#3  + #4)  -   2M_\pi^2(\bar{l}_1+2\bar{l}_2) - \frac{P(#2,#3,#4)}{#5} \right)}
\begin{align}
\tempamp{1}{s\hpr}{t\rpr}{u\rpr}{\prUN}~,\nn\\
\tempamp{2}{s\hpr}{t\hpr}{u\hpr}{\prDO}~,\nn\\
\tempamp{3}{s\rpr}{t\rpr}{u\hpr}{\prTR}~,\nn\\
\tempamp{4}{s\rpr}{t\hpr}{u\rpr}{\prCU}~.
\end{align}

\subsection{Diagrams VII}

\subsubsection{$\pi^0\pi^0 \,H\to \pi^0\pi^0$}
\renewcommand{\tempamp}[2]{T_{#1} & = -\frac{B}{F_{\pi}^4} \left\{ 2q^2+#2 + M_{\pi}^2 \frac{2q^2 - M_{\pi}^2}{#2}
\vphantom{-\frac{2Bi}{F_{\pi}^4}}
\right\} \bar{B}_0(q^2)
}

\begin{align}
\tempamp{1}{\prUN}~,\nn\\
\tempamp{2}{\prDO}~,\nn\\
\tempamp{3}{\prTR}~,\nn\\
\tempamp{4}{\prCU}~.
\end{align}

\subsubsection{$\pi^0\pi^0 \,H\to \pi^+\pi^-$}
\renewcommand{\tempamp}[3]{T_{#1} & = -\frac{B}{F_{\pi}^4} \left( #2 - M_{\pi}^2 \right)\left\{ 1 +   \frac{2q^2 - M_{\pi}^2}{#3} \right\}\bar{B}_0(q^2) 
}
\begin{align}
\tempamp{1}{s\hphantom{'}}{\prUN}~,\nn\\
\tempamp{2}{s\hphantom{'}}{\prDO}~,\nn\\
\tempamp{3}{s'}{\prTR}~,\nn\\
\tempamp{4}{s'}{\prCU}~.
\end{align}
\subsection{Diagrams VIII}
\subsubsection{$\ppnn$}
\renewcommand{\tempamp}[3]{T_{#1} & = -\frac{2B}{F_\pi^4} \left\{\frac{M_\pi^2}{2}+ \frac{\frac{3}{2}M_\pi^4 + #2^2 - 2 #2 M_\pi^2}{#3} \right\}  \bar{B}_0(#2) }

 $s$-channel diagrams:
\begin{align}
\tempamp{1}{s\hpr}{\prUN}~,\nn\\
\tempamp{2}{s\hpr}{\prDO}~,\nn\\
\tempamp{3}{s\rpr}{\prTR}~,\nn\\
\tempamp{4}{s\rpr}{\prCU}~.
\end{align}

$t$-crossed diagrams:
\begin{align}
\tempamp{1}{t\rpr}{\prUN}~,\nn\\
\tempamp{2}{t\hpr}{\prDO}~,\nn\\
\tempamp{3}{t\rpr}{\prTR}~,\nn\\
\tempamp{4}{t\hpr}{\prCU}~.
\end{align}

$u$-crossed diagrams:
\begin{align}
\tempamp{1}{u\rpr}{\prUN}~,\nn\\
\tempamp{2}{u\hpr}{\prDO}~,\nn\\
\tempamp{3}{u\hpr}{\prTR}~,\nn\\
\tempamp{4}{u\rpr}{\prCU}~.
\end{align}

\subsubsection{$\ppnc$}
In the $t$- and $u$-channel amplitudes, we define:
\begin{align}
Q(s,t,u) & = \frac{(s + u - 2M_{\pi}^2)(M_{\pi}^2 - t) + M_{\pi}^2 t}{2} \bar{B}_0(t) \nn\\
& + t (s + u - 4 M_{\pi}^2) \bar{B}_{20}(t) + 2 t (u - 2 M_{\pi}^2)\bar{B}_{21}(t)
\end{align}

$s$-channel diagrams
\renewcommand{\tempamp}[3]{T_{#1} & =  - \frac{B}{F_\pi^4}(#2 -M_\pi^2) \left( 1 + \frac{#2 +M_\pi^2}{#3} \right) \bar{B}_0(#2)}
\begin{align}
\tempamp{1}{s\hpr}{\prUN}~,\nn\\
\tempamp{2}{s\hpr}{\prDO}~,\nn\\
\tempamp{3}{s\rpr}{\prTR}~,\nn\\
\tempamp{4}{s\rpr}{\prCU}~.
\end{align}

$t$-crossed diagrams
\renewcommand{\tempamp}[5]{T_{#1} & = -\frac{2B}{F_{\pi}^4} \frac{Q(#2,#3,#4)}{#5} }
\begin{align}
\tempamp{1}{s\hpr}{t\rpr}{u\rpr}{\prUN}~,\nn\\
\tempamp{2}{s\hpr}{t\hpr}{u\hpr}{\prDO}~,\nn\\
\tempamp{3}{s\rpr}{t\rpr}{u\hpr}{\prTR}~,\nn\\
\tempamp{4}{s\rpr}{t\hpr}{u\rpr}{\prCU}~.
\end{align}

$u$-crossed diagrams:
\begin{align}
\tempamp{1}{s\hpr}{u\rpr}{t\rpr}{\prUN}~,\nn\\
\tempamp{2}{s\hpr}{u\hpr}{t\hpr}{\prDO}~,\nn\\
\tempamp{3}{s\rpr}{u\hpr}{t\rpr}{\prTR}~,\nn\\
\tempamp{4}{s\rpr}{u\rpr}{t\hpr}{\prCU}~.
\end{align}

\subsection{Diagrams IX}
In these amplitudes, the scalar source $s$ can be attached to one of the two four pion vertex (recall Fig.~\ref{fig:pipis_def}), which we denote here by the subscript $i=1,2$.
\subsubsection{$\ppnn$}
$s$-channel diagrams:
\renewcommand{\tempamp}[2]{T_{#1} & =  \frac{B}{F_\pi^4} \left(2#2+M_\pi^2 \right) \bar{B}_0(#2)}
\begin{align}
\tempamp{1}{s\hpr}~,\nn\\
\tempamp{2}{s\rpr}~.
\end{align}

$t$-crossed diagrams:
\begin{align}
\tempamp{1}{t\hpr}~,\nn\\
\tempamp{2}{t\rpr}~.
\end{align}

$u$-crossed diagrams:
\begin{align}
\tempamp{1}{u\hpr}~,\nn\\
\tempamp{2}{u\rpr}~.
\end{align}
\subsubsection{$\ppnc$}
\renewcommand{\tempamp}[2]{T_{#1} & =  \frac{B}{F_\pi^4} \left(4#2  - 3 M_\pi^2  \right)\bar{B}_0(#2)}
$s$-channel diagrams:
\begin{align}
\tempamp{1}{s\hpr}~,\nn\\
\tempamp{2}{s\rpr}~.
\end{align}
\renewcommand{\tempamp}[2]{T_{#1} & =  -\frac{B}{F_\pi^4} \left(#2  - 2 M_\pi^2  \right) \bar{B}_0(#2)}

$t$-crossed diagrams:
\begin{align}
\tempamp{1}{t\hpr}~,\nn\\
\tempamp{2}{t\rpr}~.
\end{align}

$u$-crossed diagrams:
\begin{align}
\tempamp{1}{u\hpr}~,\nn\\
\tempamp{2}{u\rpr}~.
\end{align}

\renewcommand{\tempamp}[1]{T & =
-\frac{2B}{F_{\pi}^4} \left\{ \vphantom{\frac{2B}{F_{\pi}^4}} \left(2(#1-M_{\pi}^2)(#1'-M_{\pi}^2)+M_{\pi}^4\right) {C}_0(#1,#1',q^2) + \right. \nonumber \\
&  \left. \vphantom{\frac{2B}{F_{\pi}^4}} + M_{\pi}^2 \left( \bar{B}_0(#1) + \bar{B}_0(#1') + 2 \bar{B}_0(q^2) \right)+ \frac{#1 + #1' - q^2}{2}\bar{B}_0(q^2)  \right\} }

\subsection{Diagrams X}
\subsubsection{$\ppnn$}
$s$-channel diagrams:
\begin{align}
\tempamp{s}
\end{align}

$t$-crossed diagrams:
\begin{align}
\tempamp{t}
\end{align}

$u$-crossed diagrams:
\begin{align}
\tempamp{u}
\end{align}
\subsubsection{$\ppnc$}
\renewcommand{\tempamp}[2]{T &  = -\frac{2B}{F_\pi^4} \left( \vphantom{\frac{2B}{F_\pi^4}} (#1#2-M_\pi^4) C_0(#1,#2,q^2) + (#1+#2-2M_\pi^2) \bar{B}_0(q^2) + \right. \nn\\
& \left. \vphantom{\frac{2Bi}{F_\pi^4}} + (s'-M_\pi^2)\bar{B}_0(s') + (s-M_\pi^2)\bar{B}_0(s) \right)}
$s$-channel diagrams:
\begin{align}
\tempamp{s}{s'}
\end{align}

$t$-crossed diagrams:
\renewcommand{\tempamp}[4]{T = &  -\frac{2B}{F_\pi^4} \left( \vphantom{\frac{2B}{F_\pi^4}}  \frac{(2M_\pi^2-#1)(2M_\pi^2-#2)}{2}C_0(#1,#2,q^2) +  \right. \nn \\
& \left. \vphantom{\frac{2B}{F_\pi^4}} \frac{2M_\pi^2-#1}{2}\left( \bar{B}_0(q^2) + \bar{B}_0(#1) \right) + \frac{2M_\pi^2-#2}{2}\left( \bar{B}_0(q^2) + \bar{B}_0(#2)\right) \right. \nn \\
& + \frac{#1 + #2 - q^2}{4}\bar{B}_0(q^2) + \frac{(s-s')^2-(#3-#4)^2}{2}\bar{C}_{23}(#1,#2,q^2) \nn \\
& - \left.\vphantom{\frac{2B}{F_\pi^4}} \left( s+s'-#3-#4\right) \bar{C}_{24}(#1,#2,q^2)
\right)}

\begin{align}
\tempamp{t}{t'}{u}{u'}
\end{align}

$u$-crossed diagrams:
\begin{align}
\tempamp{u}{u'}{t}{t'}
\end{align}

\subsection{Diagrams XIII}
\subsubsection{$\ppnn$}
\begin{equation}
T = \frac{B}{F_\pi^4} \left(25M_\pi^2 + 8q^2 \right)\bar{B}_0(q^2)~.
\end{equation}
\subsubsection{$\ppnc$}
\begin{equation}
T = \frac{B}{F_\pi^4} \left( 5(s+s'-M_\pi^2) + q^2 \right)\bar{B}_0(q^2)~.
\end{equation}


\end{document}